\documentclass[12pt]{article}
\usepackage{graphicx}
\usepackage{amssymb}
\usepackage{amsmath}
\usepackage{color}
\makeatletter

\@addtoreset{equation}{section}
\makeatletter

\newcommand{\D}{{\rm d}}

\newcommand{\dalm}{\kern1pt\vbox{\hrule height 0.9pt\hbox{\vrule width
0.9pt\hskip 2.5pt\vbox{\vskip 5.5pt}\hskip 3pt\vrule width 0.3pt}\hrule height
0.3pt}\kern1pt}

\def\b2hat{ {\hat b}_2 }

\begin{document}

\begin{titlepage}
\vfill
\begin{flushright}
RUP-15-14
\end{flushright}

\vfill
\begin{center}
\baselineskip=16pt
{\Large\bf 
Does the Gauss-Bonnet term stabilize wormholes?
}
\vskip 0.5cm
{\large {\sl }}
\vskip 10.mm
{\bf  Takafumi Kokubu${}^{a}$, Hideki Maeda${}^{b}$, and Tomohiro Harada$^{a}$} \\

\vskip 1cm
{
    ${}^a$ Department of Physics, Rikkyo University, Tokyo 171-8501, Japan \\
   	${}^b$ Department of Electronics and Information Engineering, Hokkai-Gakuen University, Sapporo 062-8605, Japan\\
	\texttt{takafumi-at-rikkyo.ac.jp, h-maeda-at-hgu.jp, harada-at-rikkyo.ac.jp}

     }
\vspace{6pt}
\end{center}
\vskip 0.2in
\par
\begin{center}
{\bf Abstract}
 \end{center}
\begin{quote}
The effect of the Gauss-Bonnet term on the existence and dynamical stability of thin-shell wormholes as negative tension branes is studied in the arbitrary dimensional spherically, planar, and hyperbolically symmetric spacetimes.
We consider radial perturbations against the shell for the solutions which have the Z${}_2$ symmetry and admit the general relativistic limit.
It is shown that the Gauss-Bonnet term shrinks the parameter region admitting static wormholes.
The effect of the Gauss-Bonnet term on the stability depends on the spacetime symmetry. 
For planar symmetric wormholes, the Gauss-Bonnet term does not affect their stability.
 If the coupling constant is positive but small, the Gauss-Bonnet term tends to destabilize spherically symmetric wormholes, while it stabilizes hypebolically symmetric wormholes. 
The Gauss-Bonnet term can destabilize hypebolically symmetric wormholes as a non-perturbative effect, however, spherically symmetric wormholes cannot be stable.

  \vfill
\vskip 2.mm
\end{quote}
\end{titlepage}




\tableofcontents

\newpage 

\section{Introduction}
\label{sec1}
Wormhole is a spacetime configuration connecting multiple non-timelike infinities.
Wormholes are still hypothetical objects at present but they are certainly realized as solutions for the Einstein equations.
Theoretical physicists have been attracted to traversable wormholes because they admit an apparently superluminal travel (a spacetime shortcut) as a global effect of the spacetime topology~\cite{visser,superluminal,lobo2007} and also they are available to make time machines~\cite{mt1988,mty1988,timemachine}.
(See~\cite{visser} for a standard text book and~\cite{lobo2007} for a recent review.)

As clearly seen in its Penrose diagram, the Schwarzschild vacuum spacetime with positive mass certainly possesses the wormhole structure.
The static region in this spacetime covered by the isotropic coordinates is called the Einstein-Rosen bridge connecting two distinct asymptotically flat regions~\cite{er1935}.
Of course, the Einstein-Rosen bridge is nothing but a static portion of the maximally extended Schwarzschild spacetime, so it represents just an instantaneous and non-traversable static wormhole~\cite{fw1962}.

Although the Einstein-Rosen bridge is not satisfactory as a static traversable wormhole, such a spacetime can be constructed simply by gluing two Schwarzschild exterior spacetimes~\cite{visser}.
There is no singularity in the resulting spacetime and an observer can travel from one asymptotically flat region to the other.
But unfortunately, this construction requires a massive thin shell at the junction timelike hypersurface which violates the weak energy condition.
Actually, the requirement of such exotic matters is quite generic for the wormhole construction~\cite{visser}.
From a classical viewpoint, exotic matters typically suffer from the dynamical instability.
However, wormholes might be realized by some quantum effects which effectively violate the weak energy condition.
For this reason, construction of a dynamically stable wormhole in a realistic situation has been a big challenge in gravitation physics for a long time.

In 2011, Kanti, Kleihaus, and Kunz numerically constructed four-dimensional spherically symmetric wormhole solutions in Einstein-Gauss-Bonnet-dilaton gravity and showed that they are dynamically stable against spherical perturbations~\cite{kkk2011}.
The Gauss-Bonnet term non-minimally coupled to a dilaton scalar field appears in the Lagrangian as the ghost-free quadratic correction in the low-energy limit of heterotic string theories~\cite{Gross}.
Although this Einstein-Gauss-Bonnet-dilaton theory is realized only in ten dimensions, their result gives courage and hope toward the construction of wormholes in our universe described by a well-motivated effective theory of gravity.
Then a natural question arises: Which is the main ingredient stabilizing the wormhole, the Gauss-Bonnet term or its dilaton coupling?

The main purpose of the present paper is to clarify the effect of the Gauss-Bonnet term on the dynamical stability.
For this purpose, we will study the simplest thin-shell wormhole which is made of its tension~\cite{barcelo&visser}.
While dynamical stability of thin-shell wormholes have been intensively investigated both in general relativity (Einstein gravity)~\cite{TSWs in Einstein} and in various models of modified gravity~\cite{TSWs in modified}, this is the best set-up to analyze stability as a pure gravitational effect because such a thin shell does not suffer from the matter instability.

In Einstein gravity, such thin-shell wormholes have been fully investigated by two of the present authors in the spacetimes with a maximally symmetric base manifold in arbitrary dimensions, namely arbitrary-dimensional spacetimes with spherical, planar, or hyperbolic symmetry~\cite{kh2015}.
In the vacuum case, such thin-shell wormholes are stable against radial perturbations only in the hyperbolically symmetric case with negative mass in the bulk spacetime~\cite{kh2015}.

In the present paper, we will study the same system with the Gauss-Bonnet term but without a dilaton in the Lagrangian.
Since the Gauss-Bonnet term becomes total derivative and does not affect the field equations in four or less dimensions in the absence of the non-minimal coupling to a dilaton, we will consider five or higher-dimensional spacetimes.
In comparison with the general relativistic case, the equation of motion for the shell is much more complicated.
For this reason, although thin-shell wormholes have been investigated in Einstein-Gauss-Bonnet gravity by many authors~\cite{shell-GB}, the stability analysis has not been completed yet even against radial perturbations.

The outline of the present paper is as follows.
In section II, we derive the equation of motion for a shell and the basic properties of the static shell are reviewed.
 In section III, we study the existence and stability of static thin-shell wormholes in the cases except for $k=-1$ with $m<0$.
Section IV is devoted to performing the pictorial analysis to study the same problem in the case of $k=-1$ with $m<0$.
Our results are summarized in section V.
A detailed derivation of the equation of motion for a thin shell is summarized in appendix A.
Our basic notation follows~\cite{wald}.
The convention for the Riemann curvature tensor is $[\nabla _\rho ,\nabla_\sigma]V^\mu ={R^\mu }_{\nu\rho\sigma}V^\nu$ and $R_{\mu \nu }={R^\rho }_{\mu \rho \nu }$.
The signature of the Minkowski metric is taken as diag$(-,+,+,\cdots,+,+)$, and Greek indices run over all spacetime indices.
We adopt the units in which only the $d$-dimensional gravitational constant $G_d$ is retained.

\section{Preliminaries}
\subsection{Einstein-Gauss-Bonnet gravity}
In the present paper, we consider $d(\ge 5)$-dimensional Einstein-Gauss-Bonnet gravity in vacuum, of which action is given by 
\begin{equation}
S=\frac{1}{2\kappa_d^2}\int \D^dx\sqrt{-g}\biggl(R-2\Lambda+\alpha{L}_{\rm GB}\biggl), \label{action}
\end{equation}
where $\kappa_d:=\sqrt{8\pi G_d}$ and $\Lambda$ is the cosmological constant. 
The Gauss-Bonnet term ${L}_{\rm GB}$ is defined by the following special combination of the Ricci scalar $R$, the Ricci tensor $R_{\mu\nu}$ and the Riemann tensor $R^\mu{}_{\nu\rho\sigma}$:
\begin{equation}
{L}_{\rm GB}:=R^2-4R_{\mu\nu}R^{\mu\nu}+R_{\mu\nu\rho\sigma}R^{\mu\nu\rho\sigma}.
\end{equation}
The Gauss-Bonnet term appears in the action as the ghost-free quadratic curvature correction term in the low-energy limit of heterotic superstring theory in ten dimensions (together with a dilaton)~\cite{Gross}.
In this context, the coupling constant $\alpha$ is regarded as the inverse string tension and positive definite. 
For this reason, we assume $\alpha > 0$ throughout this paper.
In addition, we put another conservative assumption $1+4{\tilde\alpha}{\tilde\Lambda}>0$, where
\begin{equation}
{\tilde\Lambda}:=\frac{2\Lambda}{(d-1)(d-2)},\qquad {\tilde\alpha}:=(d-3)(d-4)\alpha,
\end{equation}
so that the theory admits maximally symmetric vacua, namely Minkowski, de~Sitter (dS), or anti-de~Sitter (AdS) vacuum solutions.
Although there exists a maximally symmetric vacuum for $1+4{\tilde\alpha}{\tilde\Lambda}=0$, we don't consider this case for simplicity.

Variation of the action (\ref{action}) with respect to the metric $g^{\mu\nu}$ gives the following vacuum Einstein-Gauss-Bonnet equations:
\begin{equation}
{G}^\mu{}_\nu +\alpha {H}^\mu{}_\nu +\Lambda \delta^\mu{}_\nu=0, \label{beq}
\end{equation}
where
\begin{align}
{G}_{\mu\nu}:=&R_{\mu\nu}-{1\over 2}g_{\mu\nu}R, \\
{H}_{\mu\nu}:=&2\Bigl(RR_{\mu\nu}-2R_{\mu\alpha} R^\alpha{}_\nu -2R^{\alpha\beta} R_{\mu\alpha\nu\beta} + R_\mu{}^{\alpha\beta\gamma}R_{\nu\alpha\beta\gamma} \Bigr) - {1\over 2} g_{\mu\nu} {L}_{\rm GB}.
\end{align}
The tensor ${H}_{\mu\nu}$ obtained from the Gauss-Bonnet term does not give any higher-derivative term and ${H}_{\mu\nu}\equiv 0$ holds for $d\le 4$.
As a result, Einstein-Gauss-Bonnet gravity is a second-order quasi-linear theory as Einstein gravity is.

\subsection{Bulk solution}
\label{bulk}
We will study the properties of thin-shell wormholes in Einstein-Gauss-Bonnet gravity.
Such wormhole solutions are constructed by gluing two bulk solutions at a timelike hypersurface.

In the present paper, we consider the $d$-dimensional vacuum solution with a maximally symmetric base manifold~\cite{bdw} as the bulk solution, of which metric is given by 
\begin{align}
\D s_d^2=g_{\mu\nu}\D x^\mu \D x^\nu=-f(r)\D t^2+f(r)^{-1}\D r^2+r^2\gamma_{AB}\D z^A\D z^B,\label{BDW1}
\end{align}
where $z^A$ and $\gamma_{AB}~(A,B=2,3,\cdots,d-1)$ are the coordinates and the unit metric on the base manifold and 
\begin{align}
\label{BDW}
f(r) := k+\frac{r^2}{2\tilde{\alpha }}\left(1\mp \sqrt{1+\frac{4\tilde{\alpha }m}{r^{d-1}}+4{\tilde\alpha}{\tilde\Lambda}}\right). 
\end{align}
Here $k=1,0,-1$ is the curvature of the maximally symmetric base manifold corresponding to the spherical, planar, and hyperbolically symmetric spacetime, respectively.
$m$ is called the mass parameter.

The expression of the metric function (\ref{BDW}) shows that there are two branches of solutions corresponding to the different signs in front of the square root.
The branch with the minus sign, called the GR branch, allows the general relativistic limit $\alpha\to 0$ as
\begin{align}
\label{f-gr GR-limit}
f(r) = k-\frac{m}{r^{d-3}}-{\tilde\Lambda}r^2.
\end{align}
On the other hand, the metric in the branch with the plus sign, called the non-GR branch, diverges in this limit.
In the following section, we will consider the bulk solution only in the GR branch as a conservative choice.

The global structure of the spacetime (\ref{BDW1}) depending on the parameters has been completely classified~\cite{tm2005}.
There are two classes of curvature singularities in the spacetime.
One is the central curvature singularity at $r=0$.
Since we assume ${\tilde\alpha}>0$ and $1+4\tilde{\alpha }\tilde{\Lambda}>0$, the interior of  the square root becomes zero at some $r=r_{\rm b}(>0)$ for negative $m$.
This corresponds to another curvature singularity called the branch singularity and the metric becomes complex at $r<r_{\rm b}$.
In this case, the domain of the coordinate $r$ is $r\in (r_{\rm b},\infty)$.

The spacetime has a Killing horizon at $r=r_{\rm h}$ satisfying $f(r_{\rm h})=0$ depending on the parameters.
In order to construct static thin-shell wormholes, the bulk spacetime needs to be static.
For this reason, we consider the bulk solution (\ref{BDW1}) in the domain $r\in (r_{\rm h},\infty)$ if there is no branch singularity and $r\in (\max\{r_{\rm b},r_{\rm h}\},\infty)$ if there is a branch singularity.
We define the future (past) direction by increasing (decreasing) direction of $t$.

\subsection{Equation of motion for a thin shell}
A thin-shell wormhole spacetime is constructed by gluing two bulk spacetimes (\ref{BDW1}) at a timelike hypersurface $r = a$.
Here the bulk spacetimes are defined in the domain  $r \ge a(>r_{\rm h})$ and may have different parameters.
Then the junction conditions, which are the field equations (\ref{beq}) in the distributional sense, tell us the matter content on the thin shell at $r =a$.
Finally, the equation of motion for the shell is obtained as a closed system when an equation of state for the matter is assumed.

The junction condition in Einstein-Gauss-Bonnet gravity is given by 
\begin{equation} 
\label{j-condition}
[K^i_{~j}]_\pm-\delta^i_{~j}[K]_\pm+2\alpha\Bigr(3[J^i_{~j}]_\pm -\delta^i_{~j}[J]_\pm -2 P^i_{~kjl}[K^{kl}]_\pm\Bigr)=-\kappa_d^2 S^i_{~j},
\end{equation}
where $i,j=0,1,\cdots, (d-1)$ are indices for the coordinates on the timelike shell~\cite{davis2003,GB-junction}.
Here we have introduced
\begin{equation}
[X]_\pm:= X^+-X^- \, ,
\end{equation}
where $X^\pm$ is the quantity $X$ evaluated either on the $+$ or $-$ side of the shell.
In the junction conditions (\ref{j-condition}), $K^i_{~j}$ is the extrinsic curvature of the shell and $K:=h^{ij}K_{ij}$, where $h_{ij}$ is the induced metric on the shell.
Other geometrical quantities are defined by 
\begin{eqnarray}
J_{ij} &:=&{1\over 3} \left(2KK_{ik}K^{k}{}_{j}+K_{kl}K^{kl}K_{ij}-2K_{ik}K^{kl}K_{lj}-K^2 K_{ij}\right) \, , \\
P_{ikjl}&:=&{\cal R}_{ikjl}+2h_{i[l}{\cal R}_{j]k}+2h_{k[j}{\cal R}_{l]i} +{\cal R}h_{i[j}h_{l]k} \, ,
\end{eqnarray}
where ${\cal R}_{ijkl}$, ${\cal R}_{ij}$, and ${\cal R}$ are the Riemann tensor, Ricci tensor, and Ricci scalar on the shell.
$P_{ijkl}$ is the divergence-free part of the Riemann tensor ${\cal R}_{ijkl}$ satisfying $D_i P^{i}{}_{jkl}=0$, where $D_i$ is the covariant derivative on the shell.
Lastly, $S^i{}_j$ is the energy-momentum tensor on the shell, which satisfies the conservation equations $D_iS^i_{~j}=0$.

A static thin-shell wormhole is realized as a static solution for the equation of motion.
However in general, $a$ is not constant but changes in time, representing a moving shell.
In such a case, $a$ may be written as a function of the proper time $\tau$ on the shell as $a=a(\tau)$.

Now let us derive the equation of motion for the shell.
We describe the position of the thin shell as $r=a(\tau)$ and $t=T(\tau)$ in the spacetime (\ref{BDW1}) and assume the form of $S^i{}_j$ as
\begin{equation}
S^i{}_j = \mbox{diag} (-\rho,p,p,\cdots, p) + \mbox{diag}(-\sigma,-\sigma,-\sigma,\cdots, -\sigma,) \, .
\end{equation}
This assumption means that the matter on the shell consists of a perfect fluid and the constant tension $\sigma$ of the shell, where $\rho$ and $p$ are the energy density and pressure of the perfect fluid.
Assuming the Z${}_2$ symmetry for the bulk spacetime, we write down the junction conditions (\ref{j-condition}) as
\begin{align} 
\frac12\kappa_d^2(\rho+\sigma)=&-\frac{(d-2)f{\dot T}}{ a}\biggl\{1+\frac{2{\tilde\alpha}}{3}\biggl( 2\frac{{\dot a}^2}{a^2}+\frac{3k}{a^2} -\frac{f}{a^2}\biggl)\biggl\}, \label{eom1}\\
-\frac12\kappa_d^2(p-\sigma)=&-\frac{a}{f{\dot T}}\biggl\{\frac{{\ddot a}}{a}+\frac{f'}{2a}+(d-3)\biggl(\frac{{\dot a}^2}{a^2}+\frac{f}{a^2}\biggl)\biggl\}  \nonumber \\
&-\frac{2{\tilde\alpha}a}{f{\dot T}}\biggl\{\frac{d-5}{3}\biggl(\frac{{\dot a}^2}{a^2}+\frac{f}{a^2}\biggl)\biggl(2\frac{{\dot a}^2}{a^2}+\frac{3k}{a^2}-\frac{f}{a^2}\biggl) \nonumber \\
&+\biggl(2\frac{{\dot a}^2}{a^2}+\frac{k}{a^2}+\frac{f}{a^2}\biggl)\frac{{\ddot a}}{a} +\frac{f'}{2a}\biggl(\frac{k}{a^2}-\frac{f}{a^2}\biggl) \biggl\},\label{eom2}
\end{align}
where $f=f(a)$.
A dot and a prime denote the derivative with respect to $\tau$ and $a$, respectively.
(See Appendix~\ref{sec:derivation} for the details of derivation.)
The above equations give the equation of motion for a thin shell in Einstein-Gauss-Bonnet gravity.

In order to obtain the equation of motion in a closed system, an equation of state for the perfect fluid is required.
One possibility is the following linear equation of state with a constant $\gamma$ :
\begin{equation}
p=(\gamma-1)\rho. \label{eos}
\end{equation}
With this equation state, the energy-conservation equation on the shell $D_iS^i_{~j}=0$, written as
\begin{equation}
{\dot \rho}=-(d-1)(p+\rho)\frac{\dot a}{a}, \label{em-cons}
\end{equation}
is integrated to give
\begin{equation}
\rho=\frac{\rho_0}{a^{(d-1)\gamma}}, \label{energy}
\end{equation}
where $\rho_0$ is a constant.

\subsection{Effective potential for the shell}
The dynamics of the shell governed by Eqs.~(\ref{eom1}) and (\ref{eom2}) with an equation of state (\ref{eos}) can be discussed as a one-dimensional potential problem.
Then the shape of the effective potential $V(a)$ for the shell determines the stability of static configurations, namely the static wormholes.

Let us derive the effective potential $V(a)$.
Squaring Eq.~(\ref{eom1}) and using Eq.~(\ref{norm2}), we obtain
\begin{align} 
\Omega(a)^2=&\biggl(\frac{f}{a^2}+\frac{{\dot a}^2}{a^2}\biggl)\biggl\{1 +\frac{2}{3}{\tilde\alpha}\biggl( 2\frac{{\dot a}^2}{a^2}+\frac{3k}{a^2}-\frac{f}{a^2} \biggl)\biggl\}^2, \label{eom3}
\end{align}
where
\begin{align}
\Omega(a)^2:=&\frac{\kappa_d^4(\rho(a)+\sigma)^2}{4(d-2)^2}. \label{Omega01}
\end{align}
This is a cubic equation for ${\dot a}^2$.
The position of the throat $a=a_0$ for a static wormhole is obtained by solving the following algebraic equation for $a_0$:  
\begin{align} 
\Omega_0^2=\frac{f_0}{a_0^2}\biggl\{1 +\frac{2}{3}{\tilde\alpha}\biggl(\frac{3k}{a_0^2}-\frac{f_0}{a_0^2} \biggl)\biggl\}^2, \label{omegastatic}
\end{align}
where $f_0:=f(a_0)$ and $\Omega_0^2:=\Omega(a_0)^2$.

For convenience, we define
\begin{align}
A(r):=&1+\frac{4\tilde{\alpha }m}{r^{d-1}}+4{\tilde\alpha}{\tilde\Lambda}
\end{align}
with which the metric function (\ref{BDW}) in the GR branch is written as
\begin{align}
f(r) =& k+\frac{r^2}{2\tilde{\alpha }}\left(1-\sqrt{A(r)}\right). \label{metric}
\end{align}
$A>0$ is required for the real metric and the absence of branch singularity. 
In the GR branch, because of the exitence of the squre root in Eq. (\ref{metric}), the following inequality holds:
\begin{align}
r^2+2{\tilde\alpha}k-2{\tilde\alpha}f(r)>0, \label{GR branch ineq}
\end{align} 
which will be used later.

Actually, Eq.~(\ref{eom3}) admits only a single real solution for ${\dot a}^2$:
\begin{align}
{\dot a}^2=-V(a), \label{conservation law}
\end{align}
which has the form of the one-dimensional potential problem.
The effective potential $V(a)$ is defined by 
\begin{align}
V(a):=f(a)-J(a)a^2, \label{potential}
\end{align}
where $J(a)$ is defined by
\begin{align}
J(a):=&\frac{\left(B(a)-A(a)^{1/2}\right)^2}{4{\tilde\alpha}B(a)}, \label{j}\\
B(a):=&\biggl\{18{\tilde\alpha}\Omega(a)^2+A(a)^{3/2}+6\sqrt{{\tilde\alpha}\Omega(a)^2(9{\tilde\alpha}\Omega(a)^2+A(a)^{3/2})}\biggl\}^{1/3}.
\end{align}
One can see $B>A^{1/2}$.
$\Omega^2$ can be expressed in terms of $A$ and $B$ as
\begin{align}
\Omega^2=\frac{(B^3-A^{3/2})^2}{36{\tilde\alpha}B^3}. \label{Omega02}
\end{align}

\subsection{Existence conditions for static shell} 
\label{sec:existence}
Here we summarize the existence conditions for a static shell located at $a=a_0$.
Equation~(\ref{conservation law}) is interpreted as the conservation law of mechanical energy for the shell. 
By differentiating Eq.~(\ref{conservation law}) with respect to $\tau$, we obtain the equation of motion for the shell as
\begin{eqnarray}
\ddot a=-\frac{1}{2}V^\prime (a). \label{EOM}
\end{eqnarray}
From Eqs.~(\ref{conservation law}) and (\ref{EOM}), $a_0$ is determined algebraically by $V(a_0)=0$ and $V^\prime(a_0)=0$.

In addition, $a_0$ must satisfy $A(a_0)>0$ and $f(a_0)>0$. 
The latter condition $f(a_0)>0$ is called the horizon-avoidance condition in Ref.~\cite{barcelo&visser}, which simply means that the position of the throat is located in the static region of the spacetime.
Actually, this condition is always satisfied because we have $V(a_0)=0$ and Eq.~(\ref{potential}) implies $f(a)>V(a)$.

\subsection{Negative energy density of the shell} 
\label{Energy density}
In closing this section, we show that the energy density on the shell $\rho+\sigma$ must be negative for static wormholes.
The condition $\rho+\sigma\ge 0$ and Eq. (\ref{eom1}) with $a=a_0(>0)$ yields
\begin{align}
a_0^2 \le -\frac{4{\tilde\alpha}k}{2+\sqrt{A_0}}, \label{positivity01}
\end{align}
where $A_0:=A(a_0)$.
Clearly, this is not satisfied for $k=1,0$ under the assumption ${\tilde\alpha}>0$.
For $k=-1$. Eq.~(\ref{positivity01})  gives
\begin{align}
a_0^2 \le \frac{4{\tilde\alpha}}{2+\sqrt{A_0}} <2{\tilde\alpha} \label{positivity02}
\end{align}
and this is not satisfied because there is a constraint $a_0^2>2{\tilde\alpha}$ for the throat radius in the GR branch, which can be shown from the combination of Eq. (\ref{GR branch ineq}) and $f(a_0)>0$.
Now we have shown that the energy density on the shell is negative in the physical set up considered in the present paper.

\section{Static thin-shell wormholes made of pure tension} 

In the present paper, we analyze stability of the static shell located at $a=a_0$ against radial linear perturbations. 
The shape of the effective potential $V(a)$ determines the dynamical stability of the static shell, as explained below.
By Eqs.~(\ref{conservation law}) and (\ref{EOM}), $a_0$ satisfies $V(a_0)=0$ and $V^\prime(a_0)=0$.
Using $V(a_0)=V^\prime(a_0)=0$, we obtain the Taylor expansion of the potential $V(a)$ around $a=a_0$ as
\begin{equation}
V(a)=\frac{1}{2}V^{\prime \prime}(a_0)(a-a_0)^2+\mathcal O((a-a_0)^3). \label{V(a)2}
\end{equation}
The stability condition against radial perturbations for the shell is then given by
\begin{equation}
V^{\prime \prime}(a_0)>0. \label{V''>0}
\end{equation}

Hereafter we will consider the case without a perfect fluid on the shell ($\rho=p=0$) and assume $\sigma<0$.
The resulting static thin-shell wormholes are made of the pure (negative) tension $\sigma$ and satisfy the null energy condition.
This simplest set up is preferred by the minimal violation of the energy conditions.
A technical advantage in this set up is the constancy of $\Omega^2$.

\subsection{Existence conditions} 
The location of the static wormhole throat $a=a_0$ is determined by the following algebraic equation obtained by eliminating $\sigma$ from Eqs. (\ref{eom1}) and (\ref{eom2}):
\begin{align}
\biggl(\frac{f_0}{a_0}-\frac{f_0^\prime}{2}\biggl)\biggl(1+2{\tilde\alpha}\frac{k-f_0}{a_0^2}\biggl)+\frac{4{\tilde\alpha}kf_0}{a_0^3}=0, \label{static01}
\end{align}
where $f_0':=f'(a_0)$.
In the limit of $\alpha \rightarrow 0$, Eq.~(\ref{static01}) reduces to the corresponding equation in Einstein gravity~\cite{kh2015}.
The explicit form of Eq.~(\ref{static01}) is 
\begin{align}
ka_0^2\sqrt{A_0}= 2ka_0^2-\frac{(d-1)m}{2a_0^{d-5}}+4{\tilde\alpha}k^2. \label{static02}
\end{align}
This equation shows that $m=0$ is required for $k=0$ and then $a_0$ is totally undetermined in the domain where both $A_0>0$ and $f_0>0$ hold.
The metric function (\ref{BDW}) with $m=k=0$ shows that the horizon avoidance condition $f_0>0$ is satisfied only for $\Lambda<0$.
Stability of this wormhole will be investigated in Section~\ref{sec:stabilityk=0}.

On the other hand, for $k=1 ~(-1)$, the left-hand side of Eq.~(\ref{static02}) is positive (negative) and hence the throat radius must satisfy
\begin{align}
k\biggl\{2ka_0^2-\frac{(d-1)m}{2a_0^{d-5}}+4{\tilde\alpha}\biggl\}>0 \tag{$\mathcal{A}$} \label{a0-const1}
\end{align}
For $k\ne 0$, squaring Eq.~(\ref{static02}) gives the following algebraic equation for $a_0$:
\begin{align}
(3-4{\tilde\alpha}{\tilde\Lambda})a_0^{2(d-3)}+16{\tilde\alpha}ka_0^{2(d-4)}&+16{\tilde\alpha}^2a_0^{2(d-5)}\nonumber \\
&-2(d-1)kma_0^{d-3}-4d{\tilde\alpha}ma_0^{d-5}+\frac14(d-1)^2m^2=0. \label{static03}
\end{align}
Static wormhole solutions with the throat radius $a_0$ must satisfy Eq.~(\ref{static03}) and also several constraints.

The first constraint is the inequality (\ref{a0-const1}).
Another constraint comes from Eqs.~(\ref{mf-GB}) and (\ref{f^2-GB}).
Eliminating $f_0^2$, we obtain
\begin{align}
f_0=&\frac{1}{4k{\tilde\alpha}}\biggl\{\frac{(d-1)m}{a_0^{d-5}}-2(ka_0^2+2{\tilde\alpha})\biggl\}. \label{fk-1}
\end{align}
Since the right-hand side of Eq.~(\ref{fk-1}) must be positive, we have a necessary condition for physical solutions:
\begin{align}
k\biggl\{\frac{(d-1)m}{a_0^{d-5}}-2(ka_0^2+2{\tilde\alpha})\biggl\}>0. \tag{$\mathcal{B}$} \label{a0-bound}
\end{align}

We do not have to impose the condition $A_0>0$ to avoid the complex metric in the bulk spacetime because any real solution of Eq. (\ref{static03}) satisfies it.
This is shown as follows.
Eq.~(\ref{static03}) is solved for $m/a_0^{d-5}$ as
\begin{align}
\frac{m}{a_0^{d-5}}=\frac{2\left(2k(d-1)a_0^2+4d{\tilde\alpha}\pm\sqrt{(d-1)^2(1+4{\tilde\alpha}{\tilde\Lambda})a_0^4+16k(d-1){\tilde\alpha}a_0^2+16(2d-1){\tilde\alpha}^2}\right)}{(d-1)^2}. \label{y-eq}
\end{align}
Substituting this into $A_0=1+4{\tilde\alpha}{\tilde\Lambda}+4{\tilde\alpha}m/a_0^{d-1}$, we obtain
\begin{align}
A_0=&\frac{U\pm 8{\tilde\alpha}\sqrt{U-16{\tilde\alpha}^2}}{(d-1)^2a_0^4}, \label{lemma1}
\end{align}
where
\begin{align}
U:=&(d-1)^2(1+4{\tilde\alpha}{\tilde\Lambda})a_0^4+16{\tilde\alpha}\left\{k(d-1)a_0^2+2d{\tilde\alpha}\right\}.
\end{align}
For any real solution of Eq.~(\ref{static03}), the interior of the square root in Eq.~(\ref{y-eq}) is non-negative, which gives the following lower bound of ${\tilde\Lambda}$:
\begin{align}
{\tilde\Lambda}\ge &-\frac{(d-1)^2a_0^4+16(d-1){\tilde\alpha}ka_0^2+16(2d-1){\tilde\alpha}^2}{4(d-1)^2{\tilde\alpha}a_0^4}. \label{ineq-lambda}
\end{align}
This inequality implies $U\ge 16{\tilde\alpha}^2$ and hence $U$ is positive. 
Therefore $A_0$ with the plus sign in Eq.~(\ref{lemma1}) is positive.
Positivity of $A_0$ with the minus sign is shown by direct computations without using the inequality~(\ref{ineq-lambda}).

\subsection{Non-existence for $k=1$ with $m\leq0$ and $k=-1$ with $m\geq0$}
\label{sec:k1 Mplus k-1 Mminus}
It is shown that there is no static thin-shell wormhole for $k=1$ with $m\leq0$ and $k=-1$ with $m\geq0$.
For the proof, we use Eq. (\ref{conservation law}) in the following form:
\begin{align}
\left( \frac{\D \ln a}{\D \tau} \right)^2+\bar V(a)=0,
\end{align}
where
\begin{align}
\bar V(a):=\frac{f(a)}{a^2} -J(a).
\end{align}
There is no static solution if $\bar V(a)$ is monotonic.
$\bar V^\prime$ is computed to give
\begin{align}
\bar V^\prime=-\frac{1}{4{\tilde\alpha}B^2}\biggl\{\frac{8k{\tilde\alpha}B^2}{a^3}+(B^2-A)B'+BA'\biggl\}. \label{bar V deriv}
\end{align}
The following expressions;
\begin{align}
A'(a)=&-\frac{4(d-1)\tilde{\alpha }m}{a^{d}},\\
B'(a)=&\frac{A(a)^{1/2}A'(a)}{2B^2}\biggl(1+\frac{3{\tilde\alpha}\Omega^2}{\sqrt{{\tilde\alpha}\Omega^2(9{\tilde\alpha}\Omega^2+A(a)^{3/2})}}\biggl)
\end{align}
imply $B^\prime \le 0~~(\ge 0)$ and $A^\prime \le 0~~(\ge 0)$ for $m \ge 0~(m \le 0)$ with equality holding for $m=0$.
Together with the facts $A,B>0$ and $B>A^{1/2}$, it is shown that $\bar V^\prime$ is negative definite for $k=1$ with $m \le 0$ and positive definite for $k=-1$ with $m \ge 0$.

\subsection{Stability criterion} 
As explained at the beginning of this section, stability of a static thin-shell wormhole is determined by the sign of $V''(a_0)$.
In this subsection, we will derive $V''(a_0)$ in a convenient form to prove the (in)stability.

\subsubsection{General relativity} 
First let us consider Einstein gravity as a simple lesson.  
In the general relativistic limit $\alpha\to 0$, Eq.~(\ref{eom3}) reduces to 
\begin{align}
\Omega^2=&\frac{f(a)}{a^2}+\frac{{\dot a}^2}{a^2}. \label{eom3-gr0}
\end{align}
By solving Eq. (\ref{eom3-gr0}) for ${\dot a}^2$, we define a potential $V(a)$ of the conservation law of the one-dimensional potential problem. Then we directly calculate the second derivative of $V(a)$. However, without such direct calculations, in principle we can derive the form of $V''(a_0)$ by operating a systematic method below, which can be applied also in more general theories of gravity. 

Suppose we get a master equation as ${\dot a}^2+V(a)=0$.
By this master equation, ${\dot a}^2$ in Eq.~(\ref{eom3-gr0}) is replaced by $-V(a)$ to give
\begin{align}
\Omega^2=&\frac{f(a)-V(a)}{a^2}. \label{eom3-gr}
\end{align}
Differentiating this equation twice, we obtain
\begin{align}
0=&a(f'-V')-2(f-V), \label{eom3-gr2}\\
0=&a(f''-V'')-(f'-V'). \label{eom3-gr3}
\end{align}

In Einstein gravity, the metric function is 
\begin{align}
f(a)=&k-\frac{m}{a^{d-3}}-{\tilde\Lambda}a^2,\label{f-gr}
\end{align}
which satisfies
\begin{align}
f'(a)=&\frac{(d-3)(k-f)-{\tilde\Lambda}(d-1)a^2}{a},\label{df-gr}\\
f''(a)=&\frac{{\tilde\Lambda}(d-1)(d-4)a^2-(k-f)(d-2)(d-3)}{a^2}. \label{ddf-gr}
\end{align}

Substituting Eq.~(\ref{df-gr}) into Eq.~(\ref{eom3-gr2}) and evaluating it at $a=a_0$ satisfying $V(a_0)=V'(a_0)=0$, we obtain
\begin{align}
f_0(:=f(a_0))=\frac{d-3}{d-1}k-{\tilde\Lambda}a_0^2. \label{key-gr1}
\end{align}
Combining this with Eq.~(\ref{f-gr}), we obtain the algebraic equation to determine $a_0$:
\begin{align}
\frac{2k}{d-1}=&\frac{m}{a_0^{d-3}}. \label{key-gr2}
\end{align}
 For $k=0$, Eq.~(\ref{key-gr2}) requires $m=0$ and $a_0$ is totally undetermined.
For $k=1(-1)$, Eq.~(\ref{key-gr2}) requires $m>(<)0$ and the throat radius $a_0$ is given by
\begin{align}
a_0=\biggl(\frac{(d-1)m}{2k}\biggl)^{1/(d-3)}. \label{solution-GR}
\end{align}

As seen in Eq.~(\ref{solution-GR}), $\Lambda$ does not contribute to the size of the wormhole throat. However, it appears in the horizon-avoidance condition $f_0>0$.
Equation~(\ref{key-gr1}) shows that $f_0>0$ requires $\Lambda<0$ in the case of $k=0,-1$.
In the case of $\Lambda=0$,  $f_0>0$ is satisfied only for $k=1$.
In the case of $\Lambda>0$ and $k=1$, $f_0>0$ gives a constraint $a_0<a_{\rm c}^{\rm(GR)}$ on the size of the wormhole throat, where 
\begin{align}
a_{\rm c}^{\rm(GR)}:=&\biggl(\frac{(d-3)k}{(d-1){\tilde\Lambda}}\biggl)^{1/2}. \label{a-c-gr}
\end{align}
On the other hand, in the case of $\Lambda<0$ and $k=-1$, $f_0>0$ gives  $a_0>a_{\rm c}^{\rm(GR)}$.
Combining this inequality with Eq.~(\ref{solution-GR}), we obtain the range of the mass parameter admitting static wormhole solutions; $0<m<m_{\rm c}^{\rm(GR)}$ for $k=1$ with $\Lambda>0$ and $m<m_{\rm c}^{\rm(GR)}(<0)$ for $k=-1$ with $\Lambda<0$, where
\begin{align}
m_{\rm c}^{\rm(GR)}:=&\frac{2k}{d-1}\biggl(\frac{(d-3)k}{(d-1){\tilde\Lambda}}\biggl)^{(d-3)/2}. \label{m-c-gr}
\end{align}

In Einstein gravity, a simple criterion for the stability of static solutions is available. 
Substituting Eqs.~(\ref{df-gr}) and (\ref{ddf-gr}) into Eq.~(\ref{eom3-gr3}), evaluating them at $a=a_0$, we obtain
\begin{align}
V''(a_0)=&-\frac{(d-1)(d-3)m}{a_0^{d-1}}  \nonumber \\
=&-\frac{2(d-3)k}{a_0^2}, \label{ddV-gr-final}
\end{align}
where we used Eqs.~(\ref{f-gr}) and (\ref{key-gr2}).
This simple expression clearly shows that the wormhole is stable only for $k=-1$ with $m<0$~\cite{kh2015}.
Existence and stability of static thin-shell wormholes in Einstein gravity are summarized in Table~\ref{table:results-GR}.
\begin{table*}[htb]
\begin{center}
\caption{The existence and stability of Z${}_2$ symmetric static thin-shell wormholes made of pure negative tension in Einstein gravity, where $a_{\rm c}^{\rm(GR)}$ and $m_{\rm c}^{\rm(GR)}$ are defined by Eqs.~(\ref{a-c-gr}) and (\ref{m-c-gr}), respectively.  }
\label{table:results-GR}
\begin{tabular}{|c|c|c|c|c|}\hline
& & Existence  & Possible range of $a_0$ & Stability \\ \hline
$k=1$& $\Lambda > 0$ &  $0<m<m_{\rm c}^{\rm(GR)}$  & $0<a_0<a_{\rm c}^{\rm(GR)}$  & Unstable \\ \cline{2-5}
 & $\Lambda\le 0$ & $m>0$   & $a_0>0$  & Unstable \\ \hline
$k=0$ & $\Lambda \ge 0$ & None  & -- & -- \\ \cline{2-5}
& $\Lambda< 0$ & $m=0$  & $a_0>0$ & Marginally stable \\ \hline
$k=-1$& $\Lambda \ge 0$ &  None  & --  & -- \\ \cline{2-5}
 &   $\Lambda< 0$ & $m<m_{\rm c}^{\rm(GR)}(<0)$ & $a_0>a_{\rm c}^{\rm(GR)}$ & Stable \\ \hline
\end{tabular}
\end{center}
\end{table*}

\subsubsection{Einstein-Gauss-Bonnet gravity} 
Although it is more complicated, we can play this game in Einstein-Gauss-Bonnet gravity in a similar manner.
Replacing ${\dot a}^2$ by $-V(a)$ in the master equation (\ref{eom3}), we obtain
\begin{align}
\Omega^2=&\frac{f(a)-V(a)}{a^2}\biggl\{1 +\frac{2{\tilde\alpha}(-2V(a)+3k-f(a))}{3a^2}\biggl\}^2. \label{eom3-3}
\end{align}

In Einstein-Gauss-Bonnet gravity, the metric function is 
\begin{align}
f(a)=&k+\frac{a^2}{2{\tilde\alpha}}\biggl(1-\sqrt{1+4{\tilde\alpha}{\tilde\Lambda}+\frac{4{\tilde\alpha}m}{a^{d-1}}}\biggl),\label{f-GB}
\end{align}
which satisfies
\begin{align}
f'(a)=&\frac{(d-5){\tilde\alpha}(k-f)^2+(d-3)a^2(k-f)-{\tilde\Lambda}(d-1)a^4}{a\{a^2+2{\tilde\alpha}(k-f)\}},\label{df-GB}\\
f''(a)=&\frac{L(a)}{a^2\{a^2+2{\tilde\alpha}(k-f)\}^3},\label{ddf-GB}
\end{align}
where
\begin{align}
L(a):=&2(d-1)^2{\tilde\alpha}{\tilde\Lambda}^2a^8-{\tilde\Lambda}a^4(d-1)\biggl\{12{\tilde\alpha}^2(k-f)^2+12{\tilde\alpha}a^2(k-f)-(d-4)a^4\biggl\} \nonumber \\
&-(k-f)\biggl\{2(d-3)(d-5){\tilde\alpha}^3(k-f)^3+4(d^2-8d+13){\tilde\alpha}^2a^2(k-f)^2 \nonumber \\
&+3(d-2)(d-5){\tilde\alpha}a^4(k-f)+(d-2)(d-3)a^6\biggl\}.
\end{align}
Equation~(\ref{f-GB}) gives
\begin{align}
m=a^{d-3}\biggl\{-{\tilde\Lambda}a^2+(k-f(a))+{\tilde\alpha}a^{-2}(k-f(a))^2\biggl\}.\label{mf-GB}
\end{align}

Differentiating (\ref{eom3-3}) and evaluating at $a=a_0$, we obtain
\begin{align}
f_0^2=&\frac{\{(d-1)a_0^2+2k(d+1){\tilde\alpha}\}f_0+(d-1){\tilde\Lambda}a_0^4-k\{(d-3)a_0^2+(d-5){\tilde\alpha}k\}}{(d-1){\tilde\alpha}}. \label{f^2-GB}
\end{align}
where we used Eq.~(\ref{df-GB}).
This equation reduces to Eq.~(\ref{key-gr1}) for $\alpha\to 0$.
Equation~(\ref{f^2-GB}) will be used to replace $f_0^p~(p=2,3,4,\cdots)$ by $f_0$.

Differentiating Eq.~(\ref{eom3-3}) twice and using Eqs.~(\ref{df-GB}) and (\ref{ddf-GB}), we finally obtain $V''(a_0)$ in a rather compact form:
\begin{align}
V''(a_0)=&-\frac{2kP(a_0)}{a_0^2(a_0^2+2k{\tilde\alpha}+2{\tilde\alpha}f_0)(a_0^2+2k{\tilde\alpha}-2{\tilde\alpha}f_0)}, \label{ddV-gb} \\
P(a_0):= &~ 4{\tilde\alpha}^2f_0\biggl\{6k-(d-3)f_0\biggl\}+(a_0^2+2k{\tilde\alpha})\biggl\{(d-3)a_0^2+2(d-5)k{\tilde\alpha}\biggl\}, \label{defP}
\end{align}
where we have eliminated ${\tilde\Lambda}$ by using Eq.~(\ref{f^2-GB}).
This expression reduces to Eq.~(\ref{ddV-gr-final}) for $\alpha\to 0$.
Because of Eq.~(\ref{GR branch ineq}), the denominator in the expression of $V''(a_0)$ is positive and therefore the sign of the function $P(a_0)$ determines the stability of the shell.

\subsection{Effect of the Gauss-Bonnet term on the stability for ${\tilde \alpha}/a_E^2\ll 1$}
Before moving onto the full-order analysis, let us clarify how the Gauss-Bonnet term affects the stability of thin-shell wormholes in the situation where ${\tilde\alpha}$ is small.

In the general relativistic limit ${\tilde\alpha} \to 0$ with $k=\pm1$,  Eq.~(\ref{static02}) gives
\begin{align}
a_0^{d-3}=\frac{(d-1)mk}{2}=:a_{{\rm E}}^{d-3}.
\end{align}
This is the static solution in Einstein gravity which requires $mk>0$~\cite{kh2015}. 
Now we obtain the static solution for $\epsilon:={\tilde \alpha}/a_E^2\ll 1$ in a perturbative method. We expand $a_0$ in a power series of $\epsilon$ :
\begin{align}
a_0=a_{{\rm E}}+a_{(1)}\epsilon+a_{(2)}\epsilon^2+\dots. \label{taylor}
\end{align}
Substituting this expression into Eq. (\ref{static02}) and expanding it in a series of $\epsilon$, we obtain
\begin{align}
a_{(1)}=\frac{2{\tilde\Lambda}(d-1)a_{{\rm E}}^2-4(d-2)k}{(d-1)(d-3)}a_{{\rm E}}. \label{a(1)}
\end{align}
This allows us to derive the expansion of Eq. (\ref{ddV-gb}):
\begin{align}
V^{\prime\prime}(a_0)\simeq &-\frac{2(d-3)k}{a_{{\rm E}}^2}+\frac{4(d-3)ka_{(1)}\epsilon}{a_{{\rm E}}^3}+\frac{8k^2\epsilon}{a_{{\rm E}}^2} \nonumber \\
=&-\frac{2(d-3)k}{a_{{\rm E}}^2}-\frac{8k\epsilon}{a_{{\rm E}}^2}\left( \frac{d-3}{d-1}k - {\tilde\Lambda}a_{{\rm E}}^2\right). \label{ddV-GB-linear-2}
\end{align}

The first term of Eq.~(\ref{ddV-GB-linear-2}) coincides with Eq.~(\ref{ddV-gr-final}) and
inside the bracket of the second term is positive because of Eq.~(\ref{key-gr1}). 
Hence we arrive a simple conclusion about the effect of the Gauss-Bonnet term for small ${\tilde\alpha}$; it destabilizes wormholes in the spherically symmetric case ($k=1$), while it stabilizes in the hyperbolically symmetric case ($k=-1$).

\subsection{Stability for $k=0,1$} 
In this subsection, we will prove (in)stability of the static thin-shell wormhole in the framework of Einstein-Gauss-Bonnet gravity.
We are going to study the sign of $V^{\prime\prime}(a_0)$ given by Eq.~(\ref{ddV-gb}) for $k=0$ and $k=1$.
Because the analysis is much more complicated in the case of $k=-1$, it will be treated in the next section.

\subsubsection{$k=0$ with $m=0$: Marginally stable} 
\label{sec:stabilityk=0}
The analysis for $k=0$ is very simple.
For $k=0$, Eq.~(\ref{static02}) gives $m=0$ and $a_0$ is totally undetermined.
Therefore, any size of the static wormhole throat is allowed for $k=0$.
This is consistent with the fact that Eq.~(\ref{ddV-gb}) gives $V^{\prime\prime}(a_0)=0$, namely the wormhole is marginally stable.

\subsubsection{$k=1$ with $m>0$: Unstable}
\begin{figure}[htbp]
  \begin{center}
    \includegraphics[bb=0 0 516 317,width=13cm]{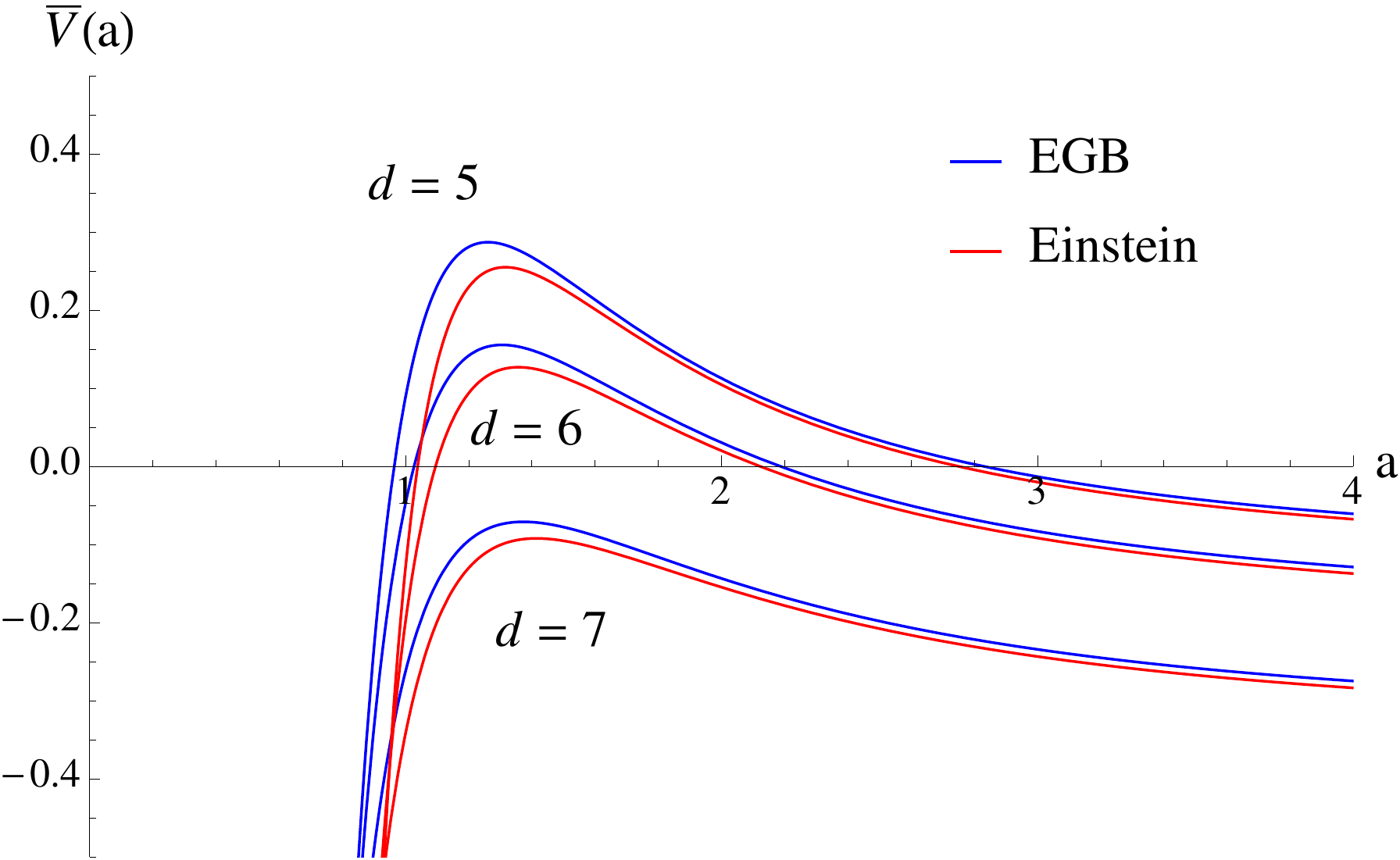}
    \caption{The potential $\bar V (a)$ for $d=5,6,7$ in Einstein and Einstein-Gauss-Bonnet (EGB) gravity with $k=1$, $\alpha=0.02$, $m=1$, $\Lambda=1$ and $\sigma=-0.1$.}
    \label{fig:EGBGR}
  \end{center}
\end{figure}

In Section~\ref{sec:k1 Mplus k-1 Mminus}, we have shown that there is no static wormhole for $k=1$ with $m\leq0$.
 In the present paper, we don't clarify the parameter region with positive $m$ admitting static wormhole solutions because they are all dynamically unstable in any case.
In Fig.~\ref{fig:EGBGR}, we plot the profile of $\bar V (a)$ with $k=1$ and $m>0$, in which there is a local maximum. This implies that the corresponding static solution is unstable.
We will prove this analytically.

For $k=1$, positivity of $P(a_0)$ in Eq.~(\ref{ddV-gb}) means instability of the static wormhole.
Using the inequality (\ref{GR branch ineq}), we evaluate the lower bound of $P(a_0)$ as
\begin{align}
P(a_0)=&4{\tilde\alpha}^2f_0\biggl\{6-(d-3)f_0\biggl\}+(a_0^2+2{\tilde\alpha})\biggl\{(d-3)a_0^2+2(d-5){\tilde\alpha}\biggl\} \nonumber \\
> &4{\tilde\alpha}^2f_0\biggl\{6-(d-3)f_0\biggl\}+2{\tilde\alpha}f_0\biggl\{(d-3)a_0^2+2(d-5){\tilde\alpha}\biggl\} \nonumber \\
= &2{\tilde\alpha}f_0\biggl\{2(d-3){\tilde\alpha}\biggl(\frac{a_0^2}{2{\tilde\alpha}}-f_0\biggl)+2(d+1){\tilde\alpha}\biggl\}  \nonumber \\
>&2{\tilde\alpha}f_0\biggl\{-2(d-3){\tilde\alpha}+2(d+1){\tilde\alpha}\biggl\}=16{\tilde\alpha}^2f_0>0.
 \label{eval-P}
\end{align}
Therefore, the wormhole is dynamically unstable.

\section{Stability for $k=-1$ with $m<0$}
In the present section, we will provide the stability analysis for $k=-1$.
Since we have shown that there is no static wormhole for $k=-1$ with $m\geq0$ in Section~\ref{sec:k1 Mplus k-1 Mminus}, we will discuss the case with $m<0$.

\subsection{Preliminaries for pictorial analysis}
For our purpose, we introduce $x:=a_0^2$ and $y:=m/a_0^{d-5}$, with which Eq. (\ref{static03}) is written as $h(x,y)=0$, where
\begin{align}
h(x,y):=&(3-4{\tilde\alpha}{\tilde\Lambda})x^2-2(d-1)kxy+\frac14(d-1)^2y^2+16{\tilde\alpha}kx-4d{\tilde\alpha}y+16{\tilde\alpha}^2. \label{def-h}
\end{align}
We adopt a pictorial analysis in the $(x,y)$ plane in the domain of $x>0$ and $y<0$.

For $3-4{\tilde\alpha}{\tilde\Lambda}\ne 0$, $h(x,y)=0$ is solved to give $x=x_{\pm}(y)$, where
\begin{align}
x_{\pm}(y):=\frac{2k\{(d-1)y-8{\tilde\alpha}\}\pm\sqrt{Z(y)}}{2(3-4{\tilde\alpha}{\tilde\Lambda})}. \label{formal sol}
\end{align}
The function $Z(y)$ in the above expression is defined by 
\begin{align}
Z(y):=&4\biggl\{(d-1)y-8{\tilde\alpha}\biggl\}^2-(3-4{\tilde\alpha}{\tilde\Lambda})\biggl\{(d-1)^2y^2-16d{\tilde\alpha}y+64{\tilde\alpha}^2\biggl\} \nonumber \\
=&(d-1)^2(1+4{\tilde\alpha}{\tilde\Lambda})y^2-16{\tilde\alpha}(4d{\tilde\alpha}{\tilde\Lambda}+d-4)y+64{\tilde\alpha}^2(1+4{\tilde\alpha}{\tilde\Lambda}). \label{def-Z}
\end{align}
In the limit $\alpha\to 0$, we obtain
\begin{align}
\lim_{\alpha\to 0}x_{+}(y)=&\frac{k(d-1)y}{2},\qquad \lim_{\alpha\to 0}x_{-}(y)=\frac{k(d-1)y}{6}.
\end{align}
Among these two, only the former satisfies Eq.~(\ref{static02}) with $\alpha=0$.
For this reason, we will focus only on $x_+(y)$ hereafter because $x_-(y)$ does not admit the general relativistic limit.

\subsubsection{Physical domain of solutions}
Solutions of Eq. (\ref{static03}) are realized as intersections of $h(x,y)=0$ with $y=m/x^{(d-5)/2}$ in the $(x,y)$ plane.
In addition, they must be located in the physical domain where all the constraints on the solutions are satisfied.  

\begin{figure}[htbp]
\begin{center}
    \includegraphics[bb=0 0 538 433,width=10cm]{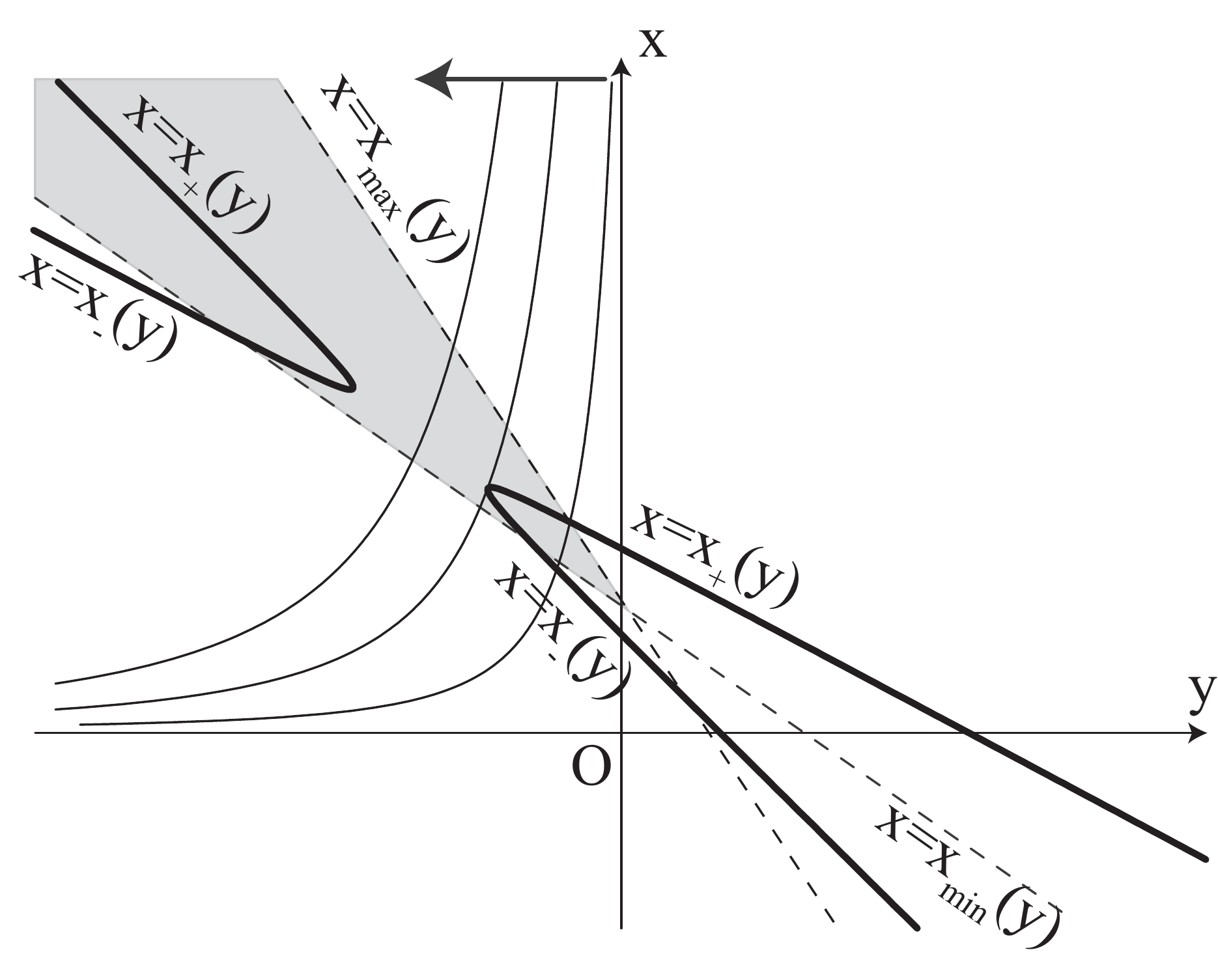}
\caption{\label{graph-Lambda3-sol} The $(x,y)$ plane for $-1<4{\tilde\alpha}{\tilde\Lambda}<-(2d-5)/(2d-1)$ with $k=-1$, $\alpha>0$, and $d=6$. The thick hyperbola consists of $x=x_{+}(y)$ and $x=x_{-}(y)$, while dashed lines consist of $x=x_{\rm max}(y)$ and $x=x_{\rm min}(y)$. Thin curves correspond to $y=m/x^{(d-5)/2}$ with three different values of negative $m$. It moves to the left as $m(<0)$ decreases. Since a static solution is realized as an intersection of the hyperbola with $y=m/x^{(d-5)/2}$ in the shadowed region, the number of static solutions depend on the value of $m$.}
\end{center}
\end{figure}

The inequality (\ref{a0-const1}) gives a constraint between $x$ and $y$ for physical solutions:
\begin{align}
x>\frac{(d-1)k}{4}y-2{\tilde\alpha}k=:x_{\rm min}(y). \label{x-min}
\end{align}
Also, the inequality (\ref{a0-bound}) gives another constraint:
\begin{align}
-2k{\tilde\alpha}<x<\frac{(d-1)k}{2}y-2k{\tilde\alpha}=:x_{\rm max}(y), \label{x-max}
\end{align}
where the left inequality comes from $a_0^2+2k{\tilde\alpha}>0$.
In summary, physical solutions must be located in the domain of $x_{\rm min}(y)<x<x_{\rm max}(y)$ and $y<0$.
Since a static solution is realized as an intersection of the hyperbola with $y=m/x^{(d-5)/2}$ in this domain, the number of static solutions depend on the value of $m$. (See Fig.~\ref{graph-Lambda3-sol} as an example.)

\subsubsection{Stable domain of solutions}

Substituting Eq.~(\ref{fk-1}) into Eq.~(\ref{defP}), we write $P(a_0)$ as a function of $x$ and $y$ as
\begin{align}
P(x,y)=&k\biggl\{(d-1)(d-3)y-16{\tilde\alpha}\biggl\}x \nonumber \\
&-\frac14\biggl\{(d-3)(d-1)^2y^2-8d(d-1){\tilde\alpha}y+128{\tilde\alpha}^2\biggl\}. \label{P}
\end{align}
The stable domain of solutions are given by $P(x,y)>0$ in the $(x,y)$ plane.
The curve $P(x,y)=0$ representing marginal stability is given by 
\begin{align}
x=\frac{(d-3)(d-1)^2y^2-8d(d-1){\tilde\alpha}y+128{\tilde\alpha}^2}{4k\{(d-1)(d-3)y-16{\tilde\alpha}\}}=:x_{\rm P}(y). \label{xp}
\end{align}
We have $x_{\rm P}(0)=-2k{\tilde\alpha}$ and 
\begin{align}
\lim_{y\to -\infty}x_{\rm P}(y)\simeq &\frac{d-1}{4k}y-\frac{2d{\tilde\alpha}}{k(d-3)}.
\end{align}

Since $P(x,y)$ is an increasing function of $x$ in the negative domain of $y$ and we have
\begin{align}
P(x_{\rm min}(y),y)=&2(d-1){\tilde\alpha}y<0,\\
P(x_{\rm max}(y),y)=&-\frac{(d-1)y\left\{8{\tilde\alpha}-(d-1)(d-3)y\right\}}{4}>0,
\end{align}
$x_{\rm P}(y)$ satisfies $x_{\rm min}(y)<x_{\rm P}(y)<x_{\rm max}(y)$.
(See Fig.~\ref{graph-Lambda0} as an example.)

Now we are ready to perform the stability analysis.
We will treat the cases of ${\Lambda} \ge 0$ and $-1<4{\tilde\alpha}{\tilde\Lambda} < 0$, separately.

\subsection{Non-existence for ${\Lambda} \ge  0$} \label{Non-existence}
In this subsection, we treat the case of ${\Lambda} \ge  0$.
Similar to the general relativistic case, Einstein-Gauss-Bonnet gravity also does not admit static thin-shell wormholes in this parameter region.

For $4{\tilde\alpha}{\tilde\Lambda}>3$, Eq.~(\ref{formal sol}) shows that $x_{+}<0<x_{-}$ holds in the domain of $y<0$ with $k=-1$.
Since $x=x_+(y)$ does not satisfy the necessary condition $x>0$, there is no static solution in this case.

In the special case of $4{\tilde\alpha}{\tilde\Lambda}=3$, $h(x,y)=0$ is solved to give
\begin{align}
x(y)=\frac{(d-1)^2y^2-16d{\tilde\alpha}y+64{\tilde\alpha}^2}{8k\{(d-1)y-8{\tilde\alpha}\}}(>0),
\end{align}
which satisfies 
\begin{align}
x_{\rm min}(y)-x(y)=\frac{(d-1)^2y^2-16(d-2){\tilde\alpha}y+64{\tilde\alpha}^2}{8k\{(d-1)y-8{\tilde\alpha}\}}>0,
\end{align}
where $x_{\rm min}(y)$ is defined in Eq.~(\ref{x-min}).
Because the necessary condition $x>x_{\rm min}$ is not satisfied, there is no static solution for $4{\tilde\alpha}{\tilde\Lambda}= 3$.

Lastly, in order to show the non-existence for $0\le 4{\tilde\alpha}{\tilde\Lambda} <3$, we will use the following fact: The sign of $x_{\rm max}(y)-x_+(y)$ is definite in some domain of $y$ if the curves $x_+(y)$ and $x_{\rm max}(y)$ do not intersect and $x_+(y)$ is continuous there.
Actually, $x_+(y)$ is continuous in the negative domain of $y$ for $4{\tilde\alpha}{\tilde\Lambda}\ge -(2d-5)/(2d-1)$ because Eq.~(\ref{def-Z}) shows that $Z(y)$ is non-negative then.

It is also possible to show that $x_+(y)$ and $x_{\rm max}(y)$ do not intersect in the negative domain of $y$.
For $\Lambda=0$, $x_+(y)=x_{\rm max}(y)$ is solved to give
\begin{align}
y=&\frac{2{\tilde\alpha}}{d-3},
\end{align}
which is positive.
For  $0< 4{\tilde\alpha}{\tilde\Lambda} <3$, the solution is 
\begin{align}
y=&\frac{4(d-1){\tilde\alpha}{\tilde\Lambda}+d-3\pm\sqrt{\left\{4(d-1){\tilde\alpha}{\tilde\Lambda}+d-3\right\}^2-4(d-1)^2{\tilde\alpha}{\tilde\Lambda}(1+4{\tilde\alpha}{\tilde\Lambda})}}{(d-1)^2{\tilde\Lambda}} \nonumber \\
=&\frac{4(d-1){\tilde\alpha}{\tilde\Lambda}+d-3\pm\sqrt{4(d-1)(d-5){\tilde\alpha}{\tilde\Lambda}+(d-3)^2}}{(d-1)^2{\tilde\Lambda}}=:y_{\rm c(\pm)}, \label{yc-pm}
\end{align}
where inside the square-root is positive for $4{\tilde\alpha}{\tilde\Lambda}>-1$.
By direct calculations, both $y_{\rm c(+)}$ and $y_{\rm c(-)}$ are shown to be positive for $\Lambda>0$.
(We note that $y_{\rm c(+)}<0$ and $y_{\rm c(-)}>0$ are satisfied for $\Lambda<0$.)

We have shown that the sign of $x_+(y)-x_{\rm max}(y)$ is definite in the domain of negative $y$ for $\Lambda>0$.
This sign is actually negative, as shown below.
From the following expression;
\begin{align}
x_{\rm max}(y)-x_+(y)=&\frac{k(d-1)(1-4{\tilde\alpha}{\tilde\Lambda})y+4k{\tilde\alpha}(1+4{\tilde\alpha}{\tilde\Lambda})-\sqrt{Z(y)}}{2(3-4{\tilde\alpha}{\tilde\Lambda})} \nonumber \\
=&\frac{-4k{\tilde\alpha}\{(d-1)y-4{\tilde\alpha}\}{\tilde\Lambda}+k(d-1)y+4k{\tilde\alpha}-\sqrt{Z(y)}}{2(3-4{\tilde\alpha}{\tilde\Lambda})}, \label{xmax-x}
\end{align}
we obtain
\begin{align}
x_{\rm max}(0)-x_+(0)=&\frac{2k{\tilde\alpha}(1+4{\tilde\alpha}{\tilde\Lambda})-4{\tilde\alpha}\sqrt{1+4{\tilde\alpha}{\tilde\Lambda}}}{3-4{\tilde\alpha}{\tilde\Lambda}},
\end{align}
which is negative for $-1<4{\tilde\alpha}{\tilde\Lambda}<3$.
Because $x_+(y)-x_{\rm max}(y)$ is continuous, it is concluded that $x_+(y)>x_{\rm max}(y)$ is satisfied in the negative domain of $y$ for $0<4{\tilde\alpha}{\tilde\Lambda}<3$.

\subsection{Pictorial analysis for $-1<4{\tilde\alpha}{\tilde\Lambda}<0$}
Now we focus on the case of $-1<4{\tilde\alpha}{\tilde\Lambda}<0$. 
In this case, there exit static solutions but their existence and stability depend on the parameters in a complicated manner.
We will clarify them by a pictorial analysis.

\subsubsection{Geometric shape of $h(x,y)=0$}

In the case of $-1<4{\tilde\alpha}{\tilde\Lambda}<0$, $h(x,y)=0$, where $h(x,y)$ is defined by Eq.~(\ref{def-h}), is a hyperbola in the $(x,y)$ plane in general.
In order to understand the shape of the hyperbola, we present
\begin{align}
h(0,y)=&\frac14(d-1)^2y^2-4d{\tilde\alpha}y+16{\tilde\alpha}^2, \\
h(x,0)=&(3-4{\tilde\alpha}{\tilde\Lambda})x^2+16{\tilde\alpha}kx+16{\tilde\alpha}^2.
\end{align}
$h(0,y)=0$ has two positive solutions $y=y_1(>0)$ and $y=y_2(>y_1)$, where 
\begin{align}
y_1:=\frac{8{\tilde\alpha}(d-\sqrt{2d-1})}{(d-1)^2},\qquad y_2:=\frac{8{\tilde\alpha}(d+\sqrt{2d-1})}{(d-1)^2}.
\end{align}
$h(x,0)=0$ also has two positive solutions $x=x_1(>0)$ and $x=x_2(>x_1)$, where 
\begin{align}
x_1:=\frac{-8{\tilde\alpha}k- 4{\tilde\alpha}\sqrt{1+4{\tilde\alpha}{\tilde\Lambda}}}{3-4{\tilde\alpha}{\tilde\Lambda}}.\qquad x_2:=&\frac{-8{\tilde\alpha}k+ 4{\tilde\alpha}\sqrt{1+4{\tilde\alpha}{\tilde\Lambda}}}{3-4{\tilde\alpha}{\tilde\Lambda}}.
\end{align}
The shape of the hyperbola drastically changes at the following critical value; 
\begin{align}
4{\tilde\alpha}{\tilde\Lambda}=-\frac{2d-5}{2d-1},
\end{align}
which is located in the domain $-1<4{\tilde\alpha}{\tilde\Lambda}<0$.
With this critical value of $\Lambda$, $h(x,y)$ becomes factored;
\begin{align}
h(x,y)=&\frac14(d-1)^2\biggl(y-\frac{64k{\tilde\alpha}x}{(d-1)\{8{\tilde\alpha}-(d-1)w\}}+\frac{64{\tilde\alpha}^2}{(d-1)^2w}\biggl) \nonumber \\
&\times \biggl(y+\frac{8kwx}{8{\tilde\alpha}-(d-1)w}+w\biggl),
\end{align}
where a constant $w$ satisfies $(d-1)^2w^2+16d{\tilde\alpha}w+64{\tilde\alpha}^2=0$, and therefore $h(x,y)=0$ consists of two straight lines.

$x=x_+(y)$ and $x=x_-(y)$ coincide when $Z(y)$ vanishes.
These points are located on $x=x_{\rm P}(y)$ only for $d=5$ or $4{\tilde\alpha}{\tilde\Lambda}=-(2d-5)/(2d-1)$.
For $4{\tilde\alpha}{\tilde\Lambda}=-(2d-5)/(2d-1)$, $Z(y)$ vanishes at
\begin{align}
y=-\frac{8{\tilde\alpha}}{d-1}=:y_0. \label{def-y0}
\end{align}
For $d=5$, it vanishes at
\begin{align}
y=\frac{{\tilde\alpha}\left\{20{\tilde\alpha}{\tilde\Lambda}+1\pm \sqrt{(4{\tilde\alpha}{\tilde\Lambda}-3)(36{\tilde\alpha}{\tilde\Lambda}+5)}\right\}}{2(1+4{\tilde\alpha}{\tilde\Lambda})}=:y_{5(\pm)}. \label{def-y5}
\end{align}
The $(x,y)$-planes for $-1<4{\tilde\alpha}{\tilde\Lambda}<0$ are drawn in Figs.~\ref{graph-Lambda1}--\ref{graph-Lambda3}.

\subsubsection{Existence of solutions}

In the present case, existence of static solutions depends on the value of the mass parameter $m$.
First of all, as in Einstein gravity, it is shown that there is no static solution for sufficiently small $|m|$.
For $-1<4{\tilde\alpha}{\tilde\Lambda}<0$, Eq.~(\ref{yc-pm}) shows that $x_+(y)<x_{\rm max}(y)$ holds in the domain $y<y_{\rm c(+)}$.
As seen in Fig.~\ref{graph-Lambda3-sol}, the curve $y=m/x^{(d-5)/2}$ moves to the right as $m(<0)$ increases approaching the $x$-axis in the limit of $m\to -0$.
Therefore, there exists a critical value $m_{\rm c}$ such that, for $m_{\rm c}<m(<0)$, the intersection of $x=x_+(y)$ with $y=m/x^{(d-5)/2}$ is located outside the physical domain in the $(x,y)$ plane and hence there is no static solution.
This critical value $m_{\rm c}$ is obtained by solving the following algebraic equations:
\begin{align}
  \begin{cases}
   2x=(d-1)ky(m_{\rm c})-4k{\tilde\alpha},&\\
    (d-1)^2{\tilde\Lambda}y(m_{\rm c})=4(d-1){\tilde\alpha}{\tilde\Lambda}+d-3+\sqrt{4(d-1)(d-5){\tilde\alpha}{\tilde\Lambda}+(d-3)^2},
  \end{cases}
\end{align}
where $y(m_{\rm c}):=m_{\rm c}/x^{(d-5)/2}$.


\begin{figure}[htbp]
\begin{center}
    \includegraphics[bb=0 0 557 408,width=10cm]{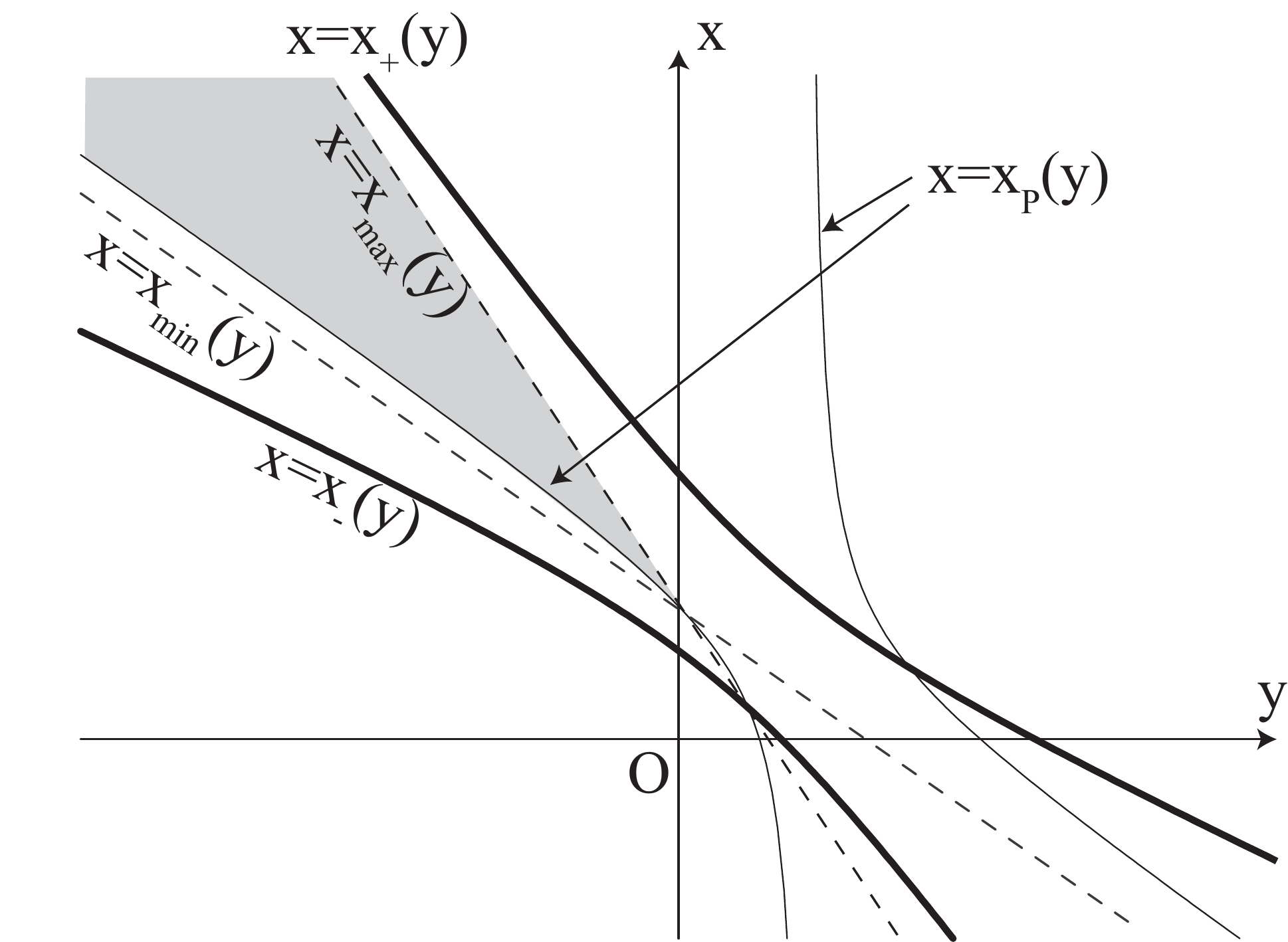}
\caption{\label{graph-Lambda0} The $(x,y)$ plane with $\Lambda=0$, ${\tilde\alpha}=1$, and $d=6$. Thick solid curves consists of $x=x_+(y)$ and $x=x_-(y)$, while thin solid curves are $x=x_{\rm P}(y)$ corresponding to $P=0$. The dashed lines consist of $x=x_{\rm max}(y)$ and $x=x_{\rm min}(y)$. The shadowed region corresponds to $P>0$ in the physical region in the domain of $x>0$, $y<0$. 
}
    \includegraphics[bb=0 0 533 408,width=10cm]{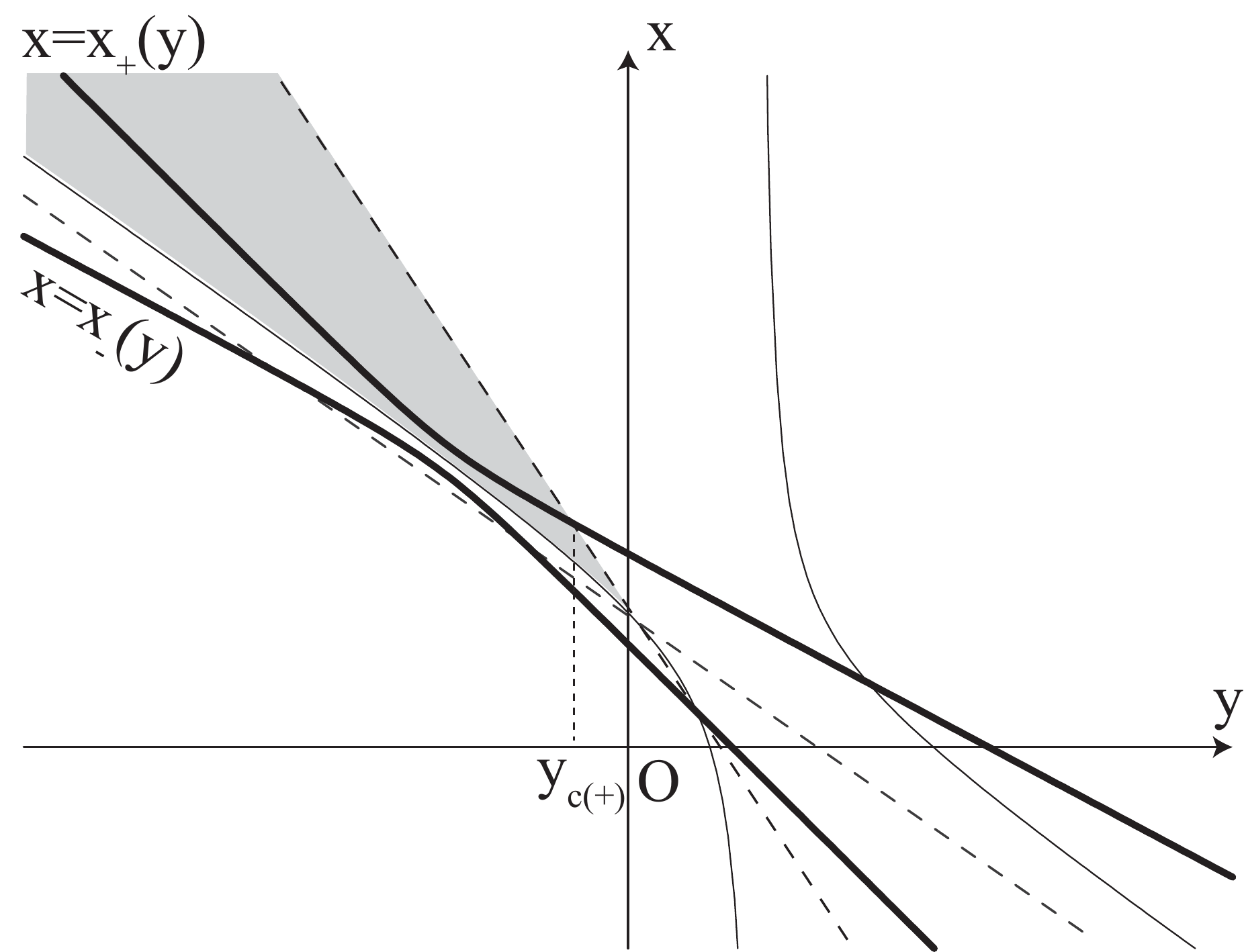}
\caption{\label{graph-Lambda1} The $(x,y)$ plane for $-(2d-5)/(2d-1)<4{\tilde\alpha}{\tilde\Lambda}<0$. If $4{\tilde\alpha}{\tilde\Lambda}$ is close to $-(2d-5)/(2d-1)$, a part of $x=x_-(y)$ enters the physical region but the corresponding solutions are unstable. }
\end{center}
\end{figure}
\begin{figure}[htbp]
\begin{center}
    \includegraphics[bb=0 0 531 419,width=10cm]{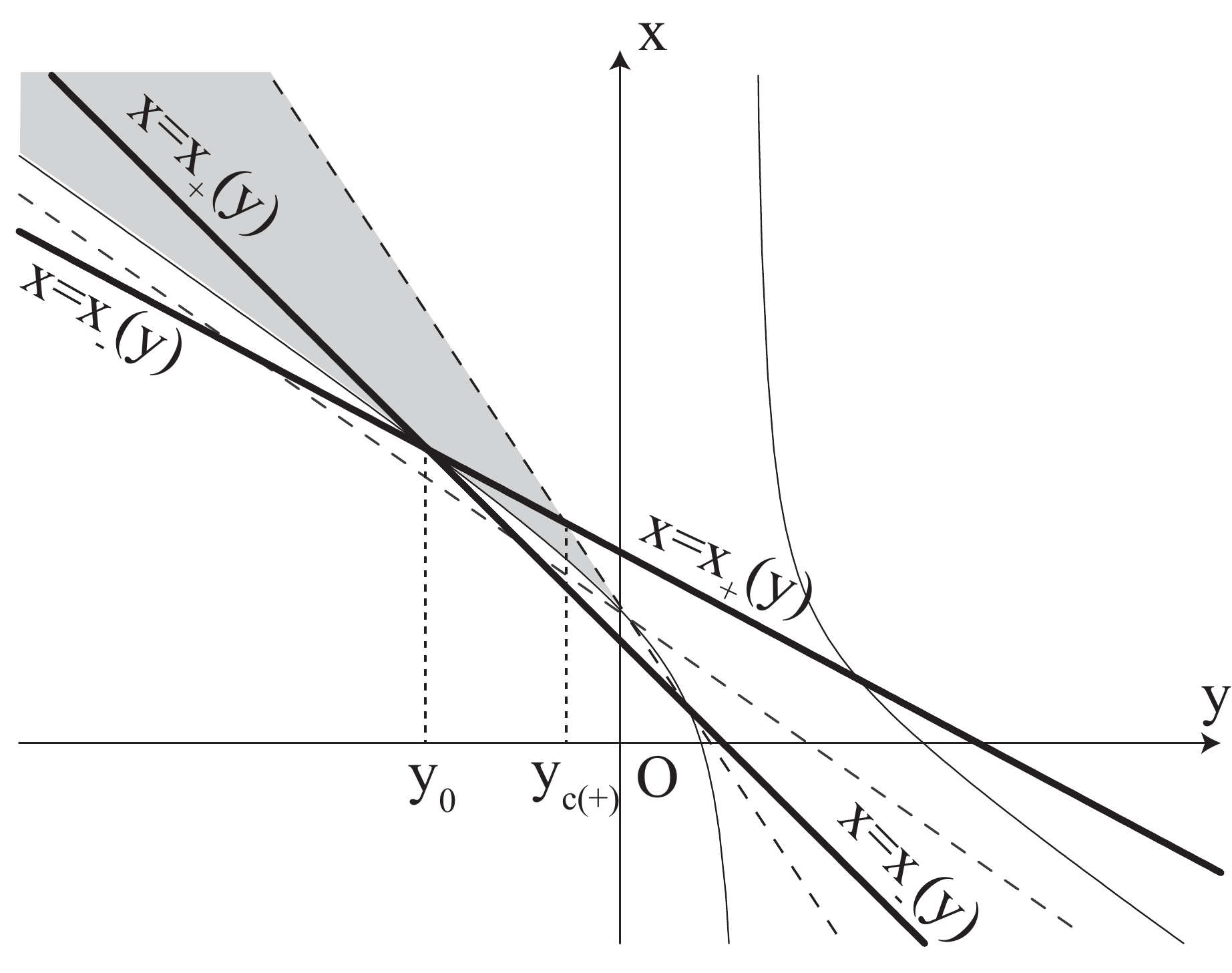}
\caption{\label{graph-Lambda2} The $(x,y)$ plane for $4{\tilde\alpha}{\tilde\Lambda}=-(2d-5)/(2d-1)$. The intersection between $x=x_+(y)$ and $x=x_-(y)$ is located on $x=x_{\rm P}(y)$ (thin solid curve) for any $d$.
}
    \includegraphics[bb=0 0 538 418,width=10cm]{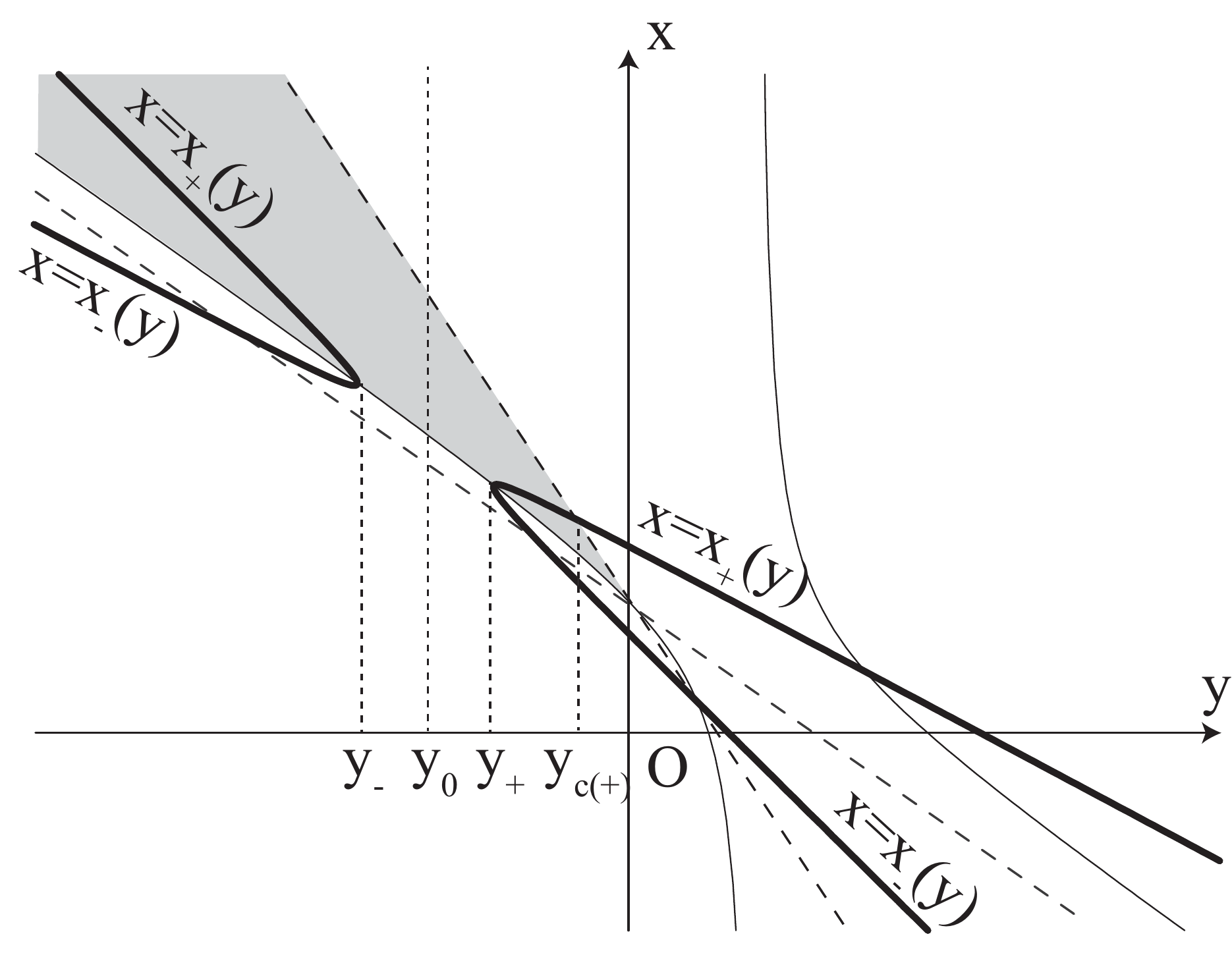}
\caption{\label{graph-Lambda3} The $(x,y)$ plane for $-1<4{\tilde\alpha}{\tilde\Lambda}<-(2d-5)/(2d-1)$. At $y=y_{\pm}$, $Z=0$ (and hence $x_+=x_-$) are satisfied. It is noted that the points $(x_+,y_+)$ and $(x_+,y_-)$ are located on $x=x_{\rm P}(y)$ (thin solid curve) only for $d=5$. }
\end{center}
\end{figure}

In the case of $-(2d-5)/(2d-1)\le 4{\tilde\alpha}{\tilde\Lambda}<0$, $Z(y)$ is non-negative and hence $x=x_+(y)$ is continuous.
Therefore, as seen in Figs.~\ref{graph-Lambda1} and \ref{graph-Lambda2}, there is a static solution for each value of $m$ satisfying $m\le m_{\rm c}$.

In the case of $-1<4{\tilde\alpha}{\tilde\Lambda}<-(2d-5)/(2d-1)$, in contrast, $Z(y)$ is negative in the domain of $y_-<y<y_+$, where
\begin{align}
y_{\pm}:=\frac{8{\tilde\alpha}\left\{4d{\tilde\alpha}{\tilde\Lambda}+(d-4)\pm \sqrt{(4{\tilde\alpha}{\tilde\Lambda}-3)[4(2d-1){\tilde\alpha}{\tilde\Lambda}+2d-5]}\right\}}{(d-1)^2(1+4{\tilde\alpha}{\tilde\Lambda})}(<0). \label{y+-}
\end{align}
As seen in Fig.~\ref{graph-Lambda3}, the curve $x=x_+(y)$ is no more continuous and does not exist in the domain of $y_-<y<y_+$.
Since the curve $y=m/x^{(d-5)/2}$ moves to the left as $m$ decreases, there exists a range of negative $m\in (m_-,m_+)$ such that the curve $y=m/x^{(d-5)/2}$ does not intersect with the hyperbola $h(x,y)=0$ and hence there is no static solution. 
This shows a sharp difference from the general relativistic case.

\subsubsection{Stability of solutions for $d=5$}

Now let us study the stability of solutions.
For this purpose, we use the following quantity:
\begin{align}
x_{+}(y)-x_{\rm P}(y) =\frac{S(y)+2k\left\{(d-1)(d-3)y-16\tilde{\alpha }\right\}\sqrt{Z(y)}}{4k(3-4 \tilde{\alpha } \tilde{\Lambda }) \left\{(d-1)(d-3) y-16 \tilde{\alpha}\right\}}, \label{stability-eq0}
\end{align}
where
\begin{align}
S(y):=&(d-3)(d-1)^2(1+4\tilde{\alpha } \tilde{\Lambda })y^2-8(d-1)\tilde{\alpha }(4d\tilde{\alpha } \tilde{\Lambda}+d-4)y+128\tilde{\alpha }^2(1+4\tilde{\alpha } \tilde{\Lambda }) \nonumber \\
=&4\tilde{\alpha } \tilde{\Lambda} \{(d-3) (d-1)^2 y^2-8 \tilde{\alpha } (d-1)dy+128 \tilde{\alpha }^2\} \nonumber \\
&~~~~~~~~~~~~~~+(d-3) (d-1)^2 y^2-8\tilde{\alpha}(d-1)(d-4) y+128\tilde{\alpha}^2.
\end{align}
If $S(y)$ is positive in some negative domain of $y$, then $x_{+}(y)>x_{\rm P}(y)$ holds there, which means that $P>0$ is satisfied at $x=x_{+}(y)$.
Therefore, if there are intersections of $x=x_{+}(y)$ with $y=m/x^{(d-5)/2}$ in the physical domain with $S(y)>0$, the corresponding static solutions are stable. 
Since the stability of static solutions is different between $d=5$ and $d\ge 6$, we treat the case of $d=5$ here and the case of $d\ge 6$ will be treated separately.

In the case of $d=5$, the solution cannot be unstable, shown as follows.
For $d=5$, $S(y)=2Z(y)$ is satisfied and so Eq.~(\ref{stability-eq0}) becomes quite simple.
Since $Z(y)\ge 0$ is satisfied on the curve $x=x_{+}(y)$, we have $x_{+}(y)\ge x_{\rm P}(y)$ there and the corresponding static solutions are stable or marginally stable.

Marginally stable static solutions are realized at $y=y_{5(\pm)}$ satisfying $Z(y_{5(\pm)})=0$, where $y_{5(\pm)}$ is given by Eq.~(\ref{def-y5}).
Because the reality of $y_{5(\pm)}$ requires $4{\tilde\alpha}{\tilde\Lambda}\le -5/9$, static solutions for $-5/9<4{\tilde\alpha}{\tilde\Lambda}<0$ are all stable. (See Fig.~\ref{graph-Lambda1}.)

On the other hand, for $-1<4{\tilde\alpha}{\tilde\Lambda}\le -5/9$, the static solutions with $m=y_{5(\pm)}$ are marginally stable, while the solutions with $m<y_{5(-)}$ or $y_{5(+)}<m$ are stable.
For $y_{5(-)}<m<y_{5(+)}$, there is no solution. 
Figure~\ref{graph-Lambda2} shows the case of $4{\tilde\alpha}{\tilde\Lambda}=-5/9$, in which $y_0$ becomes $y_{5(+)}=y_{5(-)}$ for $d=5$.
Figure~\ref{graph-Lambda3} shows the case of $-1<4{\tilde\alpha}{\tilde\Lambda}<-5/9$, in which $y_+$ and $y_-$ become $y_{5(+)}$ and $y_{5(-)}$ for $d=5$, respectively.

\subsubsection{Stability of solutions for $d\ge 6$}

In order to discuss the stability for $d\ge 6$, we evaluate the function $h(x,y)$ on the marginally stable curve $x=x_{\rm P}(y)$:
\begin{align}
h(x_{\rm P}(y),y)=&-\frac{W(y)}{16\{(d-1)(d-3)y-16\tilde{\alpha }\}^2},
\end{align}
where
\begin{align}
W(y):=&4{\tilde\alpha}{\tilde\Lambda}\left\{(d-3)(d-1)^2y^2-8d(d-1){\tilde\alpha}y+128{\tilde\alpha}^2\right\}^2 \nonumber \\
&+(d-3)^2(d-1)^4y^4-16(d-3)(d^2-5d+12)(d-1)^2{\tilde\alpha}y^3 \nonumber \\
&+64(d-1)(d^3+3d^2-52d+112){\tilde\alpha}^2y^2 \nonumber \\
&-2048(d^2-d-8){\tilde\alpha}^3y+16384{\tilde\alpha}^4.
\end{align}

We will show that the sign of $h(x_{\rm P}(y),y)$ is definite in the domain of negative $y$ for $-(2d-5)/(2d-1)<4{\tilde\alpha}{\tilde\Lambda}<0$, which means that $x=x_+(y)$ does not intersect with $x=x_{\rm P}(y)$.
Then, by continuity of the curves $x=x_+(y)$ and $x=x_{\rm P}(y)$ for $y \le 0$, the sign of $x_{+}(y)-x_{\rm P}(y)$ is the same as $x_{+}(0)-x_{\rm P}(0)$ and it is actually positive;
\begin{align}
x_{+}(0)-x_{\rm P}(0) =\frac{-2k\tilde{\alpha }(1+4\tilde{\alpha } \tilde{\Lambda })+4\tilde{\alpha}\sqrt{(1+4\tilde{\alpha } \tilde{\Lambda })}}{3-4 \tilde{\alpha } \tilde{\Lambda }}>0.
\end{align}
Therefore, $x_{+}(y)>x_{\rm P}(y)$ is satisfied for $y< 0$ and hence the corresponding static solutions are stable.
(See Fig.~\ref{graph-Lambda1}.)

In order to prove the definiteness of the sign of $h(x_{\rm P}(y),y)$ for $-(2d-5)/(2d-1)<4{\tilde\alpha}{\tilde\Lambda}<0$, we use the fact that $W(y)$ is an increasing function of $\Lambda$.
From the following two expressions;
\begin{align}
W(y)|_{{\tilde\Lambda}=0}=&(d-3)^2(d-1)^4y^4-16(d-3)(d^2-5d+12)(d-1)^2{\tilde\alpha}y^3 \nonumber \\
&+64(d-1)(d^3+3d^2-52d+112){\tilde\alpha}^2y^2 \nonumber \\
&-2048(d^2-d-8){\tilde\alpha}^3y+16384{\tilde\alpha}^4>0,\\
W(y)|_{4{\tilde\alpha}{\tilde\Lambda}=-\frac{2d-5}{2d-1}}=&\frac{4\left\{(d-1)y+8{\tilde\alpha}\right\}^2\left\{(d-1)^2(d-3)^2y^2-32d(d-3){\tilde\alpha}y+256{\tilde\alpha}^2\right\}}{2d-1}\ge 0, \label{W2}
\end{align}
it is concluded that the sign of $W(y)$ and hence the sign of $h(x_{\rm P}(y),y)$ is definite in the negative domain of $y$.

In the case of $4{\tilde\alpha}{\tilde\Lambda}= -(2d-5)/(2d-1)$, the solution can be marginally stable. 
Because the equality in Eq.~(\ref{W2}) holds only at $y=-8{\tilde\alpha}/(d-1)$, $h(x_{\rm P}(y),y)=0$ is satisfied only at $y=-8{\tilde\alpha}/(d-1)$ and the sign of  $h(x_{\rm P}(y),y)$ is definite elsewhere.
Therefore, $x_{+}(y)>x_{\rm P}(y)$ and $x_{+}(y)=x_{\rm P}(y)$ are satisfied at $y\ne -8\tilde{\alpha }/(d-1)$ and $y=-8\tilde{\alpha }/(d-1)$, respectively.
Namely, the static solution with a critical value of $m$ corresponding to $y=-8\tilde{\alpha }/(d-1)$ is marginally stable and solutions with other values of negative $m$ are stable.
(See Fig.~\ref{graph-Lambda2}.)

The situation is complicated for $-1<4{\tilde\alpha}{\tilde\Lambda}<-(2d-5)/(2d-1)$. 
In this parameter region, the solution may be dynamically unstable which shows a sharp difference from the general relativistic case.

Figure~\ref{graph-Lambda3} shows the $(x,y)$-plane in this case.
$x=x_+(y)$ exists only in the domains of $y\le y_-$ and $y_+ \le y(<0)$ because $Z(y)$ is negative in the domain of $y_-<y<y_+$, where $y_\pm$ are defined by Eq.~(\ref{y+-}) and satisfy $Z(y_\pm)=0$. 
From the following expression;
\begin{align}
Z(y_0)=&\frac{128\tilde{\alpha }^2\left\{4(2d-1){\tilde\alpha}{\tilde\Lambda}+(2d-5)\right\}}{d-1}(<0), \label{Z-crit}
\end{align}
where $y_0:=-8\tilde{\alpha }/(d-1)$, we obtain an inequality $y_-<-8\tilde{\alpha }/(d-1)<y_+$.

For our purpose, we use the fact that $S(y)$ is an increasing function of $\Lambda$.
Non-negativity of $Z$ gives the following inequality:
\begin{align}
4{\tilde\alpha}{\tilde\Lambda}\ge 3-\frac{4\{(d-1)y-8{\tilde\alpha}\}^2}{\{(d-1)^2y^2-16d{\tilde\alpha}y+64{\tilde\alpha}^2\}}. \label{lambda-bound-n}
\end{align}
This lower bound gives a lower bound of $S$:
\begin{align}
S(y)\ge &\frac{32(d-5)\tilde{\alpha }y\left\{64\tilde{\alpha }^2-(d-1)^2y^2\right\}}{(d-1)^2y^2-16d\tilde{\alpha }y+64\tilde{\alpha }^2}. \label{S-ineq}
\end{align}
If the right-hand side is positive in some domain of $y$, $x_{+}(y)>x_{\rm P}(y)$ is satisfied there and hence the corresponding static solutions are stable.

Because of the inequality $y_-<-8\tilde{\alpha }/(d-1)<y_+$, static solutions corresponding to $y\le y_-$ are dynamically stable.
In contrast, the static solution with $y=y_+$ is dynamically unstable since we have
\begin{align}
x_{+}(y_+)-x_{\rm P}(y_+)=\frac{8(d-5)\tilde{\alpha }y_+\left\{64\tilde{\alpha }^2-(d-1)^2y_+^2\right\}}{k(3-4 \tilde{\alpha } \tilde{\Lambda }) \left\{(d-1)(d-3) y_+-16 \tilde{\alpha}\right\}\{(d-1)^2y_+^2-16d\tilde{\alpha }y_++64\tilde{\alpha }^2\}}<0.
\end{align}
Figure~\ref{graph-Lambda3} shows that the dynamically unstable solutions are realized only very close to $y=y_+$ and the solutions with $y_+\ll y(<0)$ become stable.
All the results obtained in the present paper are summarized in Table~\ref{table:results}.
\begin{table*}[htb]
\begin{center}
\caption{The existence and stability of Z${}_2$ symmetric static thin-shell wormholes made of pure negative tension in the GR branch with ${\tilde\alpha}>0$ and $1+4{\tilde\alpha}{\tilde\Lambda}>0$. "S", "M", "U" stand for "Stable", "Marginally stable", and "Unstable", respectively }
\label{table:results}
\begin{tabular}{|c|c|c|c|}\hline
& & Static solutions exist?  & Stability \\ \hline
$k=1$& $m>0$ & Yes  & U \\ \cline{2-4}
& $m \le 0$ & No & -- \\ \hline
$k=0$ & $m=0$ & $\Lambda\ge 0$: No   & -- \\ \cline{3-4}
&          &  $\Lambda < 0$: Yes  & M  \\ \cline{2-4}
& $m \ne 0$ & No  & -- \\ \hline
$k=-1$& $m \ge 0$ &  No  & --  \\ \cline{2-4}
 & $m<0$ &  ${\Lambda} \ge 0$ : No  & --  \\ \cline{3-4}
&          &  $-(2d-5)/(2d-1) < 4{\tilde\alpha}{\tilde\Lambda} < 0$: Yes  & S  \\ \cline{3-4}
&          &  $4{\tilde\alpha}{\tilde\Lambda}=-(2d-5)/(2d-1)$: Yes  & S or M \\ \cline{3-4}
 &         & $-1< 4{\tilde\alpha}{\tilde\Lambda} < -(2d-5)/(2d-1)$ with $d=5$: Yes & S or M \\ \cline{3-4}
 &         & $-1< 4{\tilde\alpha}{\tilde\Lambda} < -(2d-5)/(2d-1)$ with $d\ge 6$: Yes & S, M, or U \\ \hline
\end{tabular}
\end{center}
\end{table*}

\section{Summary and discussions}
In the present paper, $d(\ge 5)$-dimensional static thin-shell wormholes with the Z${}_2$ symmetry have been investigated in the spherically ($k=1$), planar ($k=0$), or hyperbolically ($k=-1$) symmetric spacetime in Einstein-Gauss-Bonnet gravity.
For our primary motivation to reveal the effect of the Gauss-Bonnet term on the static configuration and dynamical stability of a wormhole, we have studied the stability against linear perturbations preserving symmetries in the simplest set up where the thin shell is made of pure negative tension, which satisfies the null energy condition. 

In this system, the dynamics of the shell can be treated as a one-dimensional potential problem characterized by a mass parameter $m$ in the vacuum bulk spacetime for a given value of $d$, $k$, the cosmological constant $\Lambda$, and the Gauss-Bonnet coupling constant $\alpha$.
We have studied solutions which admit the general relativistic limit $\alpha\to 0$ and considered a very conservative region in the parameter space.
The shape of the effective potential for the shell dynamics clarifies possible static configurations of a wormhole and their dynamical stability.

As seen in Tables~\ref{table:results-GR} and \ref{table:results}, the results with and without the Gauss-Bonnet term are similar in many cases.
For $k=1$, static wormholes require $m>0$ and they are dynamically unstable.
For $k=0$, static wormholes require $m=0$ and $\Lambda<0$ and they are marginally stable.
For $k=-1$, $m<0$ and $\Lambda<0$ are required for static wormholes.

We have clarified the effect of the Gauss-Bonnet term on the stability in a perturbative method by expanding the equation in a power series of $\tilde \alpha$.
We have shown that, for ${\tilde \alpha}/a_E^2\ll 1$, the Gauss-Bonnet term tends to destabilize spherically symmetric thin-shell wormholes ($k=1$), while it stabilizes hyperbolically symmetric wormholes ($k=-1$). 
For planar symmetric wormholes ($k=0$), the Gauss-Bonnet term does not affect their stability and they are marginally stable, same as in Einstein gravity. 
However, we have observed that the non-perturbative effect is quite non-trivial. 

Notable difference between Einstein gravity and Einstein-Gauss-Bonnet gravity appears in the case of $k=-1$.
In Einstein gravity, static wormholes exist when the mass parameter $m$ is less than a critical negative value and they are dynamically stable.
This is also the case in Einstein-Gauss-Bonnet gravity for $-(2d-5)/(2d-1) < 4{\tilde\alpha}{\tilde\Lambda} < 0$.
However, for $4{\tilde\alpha}{\tilde\Lambda}=-(2d-5)/(2d-1)$, a static wormhole becomes marginally stable if $m$ is fine-tuned.
For $-1<4{\tilde\alpha}{\tilde\Lambda}<-(2d-5)/(2d-1)$, in contrast, static wormholes cease to exist for a finite range of $m$ and furthermore, dynamical property of the wormhole is different for $d=5$ and $d\ge 6$.
For $d=5$, static wormholes are generically stable but become marginally stable if $m$ is fine-tuned.
For $d\ge 6$, in addition to them, wormholes are dynamically unstable in a finite range of $m$.
In summary for $k=-1$, the Gauss-Bonnet term shrinks the parameter region admitting static wormholes and tends to destabilize them non-perturbatively.

As the effect of the Gauss-Bonnet term on the existence and stability of static wormholes has been revealed in the present paper, the effect of its dilaton coupling is now of great interest.
Unfortunately in the presence of a dilaton, exact bulk solutions are not available to construct thin-shell wormholes.
Nevertheless, this is a promising direction of research leading to understand the result in~\cite{kkk2011}.
We hope that the result will be reported elsewhere.

\subsection*{Acknowledgements}
The author thanks Tsutomu Kobayashi, Shuichiro Yokoyama, Takahisa Igata and Mandar Patil for useful comments and discussions. 
T.H. was partially supported by the Grant-in-Aid No. 26400282 for Scientific
Research Fund of the Ministry of Education, Culture, Sports, Science and Technology, Japan.

\appendix

\section{Derivation of the equation of motion for a thin shell}
\label{sec:derivation}
In this appendix, we present the details how to derive the equation of motion for the shell (\ref{eom1}) and (\ref{eom2}) from the junction conditions (\ref{j-condition}).

For the following vacuum bulk metric (\ref{BDW1});
\begin{align}
\D s_d^2=&g_{\mu\nu}\D x^\mu \D x^\nu=-f(r)\D t^2+f(r)^{-1}\D r^2+r^2\gamma_{AB}\D z^A\D z^B, \\
f(r) :=& k+\frac{r^2}{2\tilde{\alpha }}\left(1\mp \sqrt{1+\frac{4\tilde{\alpha }m}{r^{d-1}}+4{\tilde\alpha}{\tilde\Lambda}}\right),
\end{align}
the non-vanishing components of the Levi-Civit\'a connection are given by 
\begin{align}
\begin{aligned}
\Gamma^r_{~tt}&=\frac12 f\frac{\D f}{\D r},\quad \Gamma^t_{~tr}=\frac{1}{2f}\frac{\D f}{\D r}, \quad \Gamma^r_{~rr}=-\frac{1}{2f}\frac{\D f}{\D r}, \\
{\Gamma ^r}_{AB}&=-r f \gamma _{AB},\quad 
{\Gamma ^A}_{Br}=\frac{1}{r}{\delta ^A}_B, \quad {\Gamma ^A}_{BC}={\hat{\Gamma} ^A}_{~BC}(z),
\end{aligned}
\end{align}
where ${\hat \Gamma ^A}_{~BC}$ is the Levi-Civit\'a connection on the maximally symmetric base manifold.

In this spacetime, the position of the thin shell is described by $r=a(\tau)$ and $t=T(\tau)$, where $\tau$ is the proper time on the shell.
The future directed unit tangent vector to the shell is 
\begin{equation}
u^\mu\frac{\partial}{\partial x^\mu}={\dot T} \frac{\partial}{\partial t}+{\dot a}\frac{\partial}{\partial r}\, ,
\end{equation}
of which normalization condition $u_\mu u^\mu=-1$ is written as
\begin{equation}
1=f(a){\dot T}^2-\frac{{\dot a}^2}{f(a)}\, , \label{norm}
\end{equation}
where a dot denotes the differentiation with respect to $\tau$.
The unit normal one-form $n_\mu$ to the shell is given by
\begin{equation}
n_\mu \D x^\mu =-{\dot a}\D t+{\dot T}\D r\, ,
\end{equation}
which satisfies $n_\mu u^\mu=0$ and $n_\mu n^\mu=1$.
The vector $n^\mu(\partial/\partial x^\mu)$ is pointing increasing direction of $r$. 

The $(d-1)$-dimensional induced metric $h_{ij}$ on the shell is given by 
\begin{equation}
\label{1stout} \D s_{d-1}^2=h_{ij}(\xi)\D \xi^i\D \xi^j= -\D \tau^2+a(\tau)^2\gamma_{AB}\D z^A\D z^B\, .
\end{equation}
where $\xi^0=\tau$.
Non-zero components of the Levi-Civit\'a connection ${}^{(d-1)}{\Gamma}{}^i_{jk}$ in this spacetime are 
\begin{eqnarray}
{}^{(d-1)}{\Gamma}{}^\tau_{AB}=a{\dot a}\gamma_{AB}, \quad {}^{(d-1)}{\Gamma}{}^A_{\tau B}=\frac{\dot a}{a}\delta^A_B, \quad {}^{(d-1)}{\Gamma}{}^A_{BC}={\hat \Gamma}^A_{BC}.
\end{eqnarray}  
From our definition of the Riemann tensor;
\begin{eqnarray}
{R}^\mu_{~\nu\rho\sigma}=\partial_\rho\Gamma^\mu_{~\nu\sigma}-\partial_\sigma\Gamma^\mu_{~\nu\rho} +\Gamma^\mu_{~\kappa\rho}\Gamma^\kappa_{~\nu\sigma}-\Gamma^\mu_{~\kappa\sigma}\Gamma^\kappa_{~\nu\rho},
\end{eqnarray}
the non-zero components of the Riemann tensor ${\cal R}^i_{~ijk}$, Ricci tensor ${\cal R}_{ij}$, and Ricci scalar ${\cal R}$ are computed to give
\begin{align}
{\cal R}^\tau_{~B\tau D}=&a{\ddot a}\gamma_{BD},\qquad {\cal R}^A_{~BCD}=(k +{\dot a}^2)(\delta^A_{~C}\gamma_{BD}-\delta^A_{~D}\gamma_{BC}), \\
{\cal R}{}_{\tau\tau}=&-(d-2)\frac{{\ddot a}}{a},\qquad {\cal R}{}_{AB}=\biggl\{a{\ddot a}+(d-3)(k +{\dot a}^2)\biggl\}\gamma_{AB},\\
{\cal R}=&2(d-2)\frac{{\ddot a}}{a}+(d-2)(d-3)\biggl(\frac{k}{a^2} +\frac{{\dot a}^2}{a^2}\biggl).
\end{align} 
From these expressions, we obtain the non-zero components of $P^i_{~jkl}$:
\begin{align}
P^\tau_{~B\tau D}=&\frac12(d-3)(d-4)(k +{\dot a}^2)\gamma_{BD},\\
P^A_{~BCD}=&(d-4)\biggl\{a{\ddot a}+\frac12(d-5)(k+{\dot a}^2)\biggl\}(\delta^A_{~C}\gamma_{BD}-\delta^A_{~D}\gamma_{BC}).
\end{align}

The extrinsic curvature of the shell is computed from the following definition:
\begin{align}
K_{ij}:=&(\nabla_\mu n_{\nu})e^\mu_i e^\nu_j \nonumber \\
=&-n_{\mu}e_{i,j}^\mu-\Gamma^\kappa_{\mu\nu}n_\kappa e_i^\mu e_j^\nu \, ,
\end{align}
where $e^\mu_i := \partial x^\mu/\partial \xi^i$. 
Using 
\begin{equation}
e^0_i\D \xi^i ={\dot T}\D \tau \, , \qquad e^1_i\D \xi^i = {\dot a}\D \tau \, , \qquad
e^A_i\D \xi^i = \delta^A_B \D z^B \, ,
\end{equation}
and Eq.~(\ref{norm}) together with its derivative with respect to $\tau$, we obtain the non-zero components of $K^i{}_j$:
\begin{equation}
K^\tau{}_\tau = \frac{1}{f{\dot T}}\left({\ddot a}+\frac{f'}{2}\right)\, , \qquad K^A{}_B = \frac{f{\dot T}}{ a}\delta^A{}_B \, ,
\end{equation}
where a prime denotes the derivative with respect to $a$.
From the above expressions, we compute
\begin{align}
K=& \frac{1}{f{\dot T}}\left({\ddot a}+\frac{f'}{2}\right)+\frac{(d-2)f{\dot T}}{ a},\\
K_{ij}K^{ij}=&  \frac{1}{f^2{\dot T}^2}\left({\ddot a}+\frac{f'}{2}\right)^2+(d-2)\biggl(\frac{f{\dot T}}{ a}\biggl)^2,\\
J^\tau{}_\tau =&-\frac{(d-2)(d-3)}{3}\frac{f{\dot T}}{a^2}\left({\ddot a}+\frac{f'}{2}\right)\, , \\
J^A{}_B =&-\frac{(d-3)f{\dot T}}{3a}\biggl\{\frac{2}{ a}\left({\ddot a}+\frac{f'}{2}\right)+(d-4)\biggl(\frac{f{\dot T}}{ a}\biggl)^2\biggl\} \delta^A{}_B \, , \\
J=&-\frac{(d-2)(d-3)f{\dot T}}{3a}\biggl\{\frac{3}{ a}\left({\ddot a}+\frac{f'}{2}\right)+(d-4)\biggl(\frac{f{\dot T}}{ a}\biggl)^2\biggl\}.
\end{align}

Now we are ready to write down the equation of motion for the shell.
Under the assumptions of the Z${}_2$ symmetry and the form of $S^i{}_j$ as
\begin{equation}
S^i{}_j = \mbox{diag} (-\rho,p,p,\cdots, p) + \mbox{diag}(-\sigma,-\sigma,-\sigma,\cdots, -\sigma,) \, ,
\end{equation}
the junction conditions (\ref{j-condition}) give (\ref{eom1}) and (\ref{eom2}), where we used Eq.~(\ref{norm}) in the following form:
\begin{equation}
\biggl(\frac{f{\dot T}}{ a}\biggl)^2=\frac{f}{a^2}+\frac{{\dot a}^2}{a^2}. \label{norm2}
\end{equation}

\end{document}